\documentclass[11pt, a4paper]{article}
\usepackage{graphicx} 
\usepackage[left = 2.5 cm,right = 2.5 cm, top = 2.5 cm,bottom = 2 cm]{geometry}
\usepackage{bm}
\usepackage{xcolor}
\usepackage{bigints}
\usepackage{algorithm}
\usepackage{algpseudocode}
\usepackage{hyperref}
\usepackage{float}
\usepackage{amsmath, amsfonts, amssymb}
\usepackage{graphicx}
\usepackage{subcaption}

\usepackage[backend=biber, sorting=none]{biblatex} 
\renewcommand{\rm}[1]{\mathrm{#1}}

\newcommand{\NFP}{n_{\rm{FP}}}

\newcommand{\nr}{N_{\rho}}
\newcommand{\nt}{N_{\theta}}
\newcommand{\nz}{N_{\zeta}}

\algrenewcommand\alglinenumber[1]{\normalsize\bfseries #1:}

\title{AGNI: A differentiable MHD stability solver \& optimizer for magnetic confinement fusion devices}
\author{Rahul Gaur$^{1, *}$, Sanket Patil$^{1}$, Prateek Gupta$^{2}$, Djin Patch$^{1}$, Tony Qian$^{1}$ \\[6pt]
\small $^{1}$University of Wisconsin--Madison, Madison, Wisconsin, USA\\
\small $^{2}$Indian Institute of Technology-Delhi, New Delhi, India\\[4pt]
\small $^{*}$Correspondence:
\texttt{rgaur@wisc.edu, rgaur@terpmail.umd.edu}
}
\vspace{10pt}

\date{August 2026}

\begin{document}

\maketitle

\begin{abstract}
   The existence of an ideal MagnetoHydroDynamic (MHD) equilibrium does not guarantee its stability. Finite toroidal mode number ($n$) instabilities degrade performance in both tokamaks and stellarators and differentiable stability optimization tools to date have operated only in the infinite-$n$ limit. We present~\texttt{AGNI} (Analysis of Global Normal modes in Ideal MHD), a GPU-accelerated, automatically differentiable finite-$n$ ideal MHD stability solver and optimizer.~\texttt{AGNI} discretizes the ideal MHD energy principle pseudospectrally in real space using differentiation matrices and geometric coefficients from a~\texttt{DESC} equilibrium, giving a variational eigenvalue problem for the plasma displacement, and efficiently finds the most unstable modes. Built on~\texttt{jax},~\texttt{AGNI} yields reverse-mode gradients of the growth rate with respect to boundary-shape and profile parameters without re-solving the equilibrium. We benchmark~\texttt{AGNI} against the initial-value code~\texttt{NIMSTELL} for a modified Landreman–Buller–Drevlak quasi-helically symmetric stellarator and a DSHAPE tokamak, recovering the dominant mode with agreement in terms of growth rate and eigenfunction structure, and verify the automatic differentiation gradients against central finite differences. We quantify CPU and GPU cost for eigenvalue and gradient evaluation, establish the finite-precision limit on resolving near-marginal eigenvalues, and present a numerical scheme to impose incompressibility, compatible with gradient-based optimization.~\texttt{AGNI} will allow us to optimize tokamaks, stellarators, and mirrors against ideal MHD instabilities.
\end{abstract}

\section{Introduction}
Magnetic confinement fusion is the most promising path to achieving self-sustained nuclear fusion. Devices such as tokamaks, stellarators, and mirrors generate energy by confining a hot plasma that is hundreds of millions of degrees. The equilibrium state of the hot, magnetically confined plasma is often calculated and described using the ideal MagnetoHydroDynamic (MHD) model. 

However, the existence of an ideal MHD equilibrium does not guarantee its stability. In tokamaks, ideal MHD instability is often directly related to the occurrence of disruptions~\cite{snyder2004nf}. In stellarators, even though these instabilities do not cause disruptions, they can significantly degrade performance and depend on the topology of the stellarator magnetic field~\cite{Wright2024NonlinearMHDQA, pandya2015, weller2006}. 
Ideal MHD stability is categorized by the periodicity of the unstable mode in the toroidal direction --- the toroidal mode number $n$. In general, small $n$ corresponds to large scale instabilities whereas large $n$ corresponds to small scale instabilities. Until now, we have only had differentiable stability optimization tools that work in the infinite-$n$ ($n \rightarrow \infty$) limit.
Therefore, it would be advantageous to have a reactor design tool to solve for and optimize against finite-$n$ ideal MHD instability.

The computational study of ideal MHD stability has a long lineage, almost all of it built on solving the ideal MHD energy principle~\cite{bernsteinenergyprinciple} as an eigenvalue problem where the magnitude of the eigenvalue is proportional to the square of the growth rate and the sign determines whether the equilibrium is stable or unstable. In the late 1970s and early 1980s, variational eigenvalue codes~\texttt{PEST}~\cite{Grimm1976PEST},~\texttt{ERATO}~\cite{Gruber1981ERATO}, and~\texttt{GATO}~\cite{Bernard1981GATO} established the standard approach: discretize the potential energy $\delta W$ over a 2D axisymmetric domain and solve the resulting matrix eigenvalue problem for low-$n$ global modes. The finite-element and ``finite-hybrid-element'' formulation that underpins~\texttt{ERATO}, together with the elimination of spectral pollution, was developed in detail by Gruber and Rappaz~\cite{Gruber1985FiniteElementMHD}. These codes remain the conceptual basis for nearly every modern stability solver.

Two needs drove the field forward. First, the tokamak edge demanded intermediate-to-high-$n$ capability:~\texttt{GATO} and similar low-$n$ codes could not reach the $n \approx 3$-$20$ peeling-ballooning modes that limit the pedestal and trigger edge-localized modes, motivating the successor code~\texttt{ELITE}~\cite{Wilson2002ELITE}, which expands in inverse $n$ and resolves the full edge spectrum. Related codes such as~\texttt{MARS}~\cite{Bondeson1992MARS} added plasma rotation and resistive-wall physics along with the ability to solve the perturbed force balance problem instead of the variational eigenvalue problem. To quantify the effect of resonant magnetic fluctuations in a tokamak, codes such as~\texttt{GPEC}~\cite{park2007computation} calculated the perturbed tokamak equilibrium and its stability using the~\texttt{DCON}~\cite{glasser2016direct} algorithm. Second, stellarators required fully 3D treatment, delivered by~\texttt{TERPSICHORE}~\cite{Anderson1990TERPSICHORE} and~\texttt{CAS3D}~\cite{Nuhrenberg1996CAS3D}, both of which take a~\texttt{VMEC}~\cite{hirshman1983steepest_VMEC} equilibrium, expand the potential energy integral $\delta W$ using well-defined basis functions, and compute the 3D eigenspectra using efficient sparse matrix algorithms~\cite{anderson_methods_1990, chen1998shift}. More recently, the 3D Newcomb-criterion code~\texttt{DCON3D}~\cite{glasser_dcon3d_2025} has extended the fast~\texttt{DCON} algorithm based on Newcomb's~\cite{newcombcriterion} criterion to a stellarator. However, building a 3D stability solver and optimizer, and optimizing equilibria with a robust, easily accessible finite-$n$ stability solver, is still an open problem. 

This work presents the~\texttt{AGNI} (Analysis of Global Normal modes in Ideal MHD) code, which, to our knowledge, is the first differentiable, GPU-accelerated, finite-$n$ ideal MHD stability solver and optimizer intended to span multiple confinement concepts. The eventual scope includes tokamaks, stellarators, and magnetic mirrors but here we focus on stellarators and tokamaks. Using the~\texttt{DESC}~\cite{dudt2020desc, panici2023desc, conlin2023desc} equilibrium code along with the~\texttt{jax}~\cite{jax2018github} package, the solver calculates an automatically differentiable energy integral which allows us to make growth rates directly optimizable, in both the low- and high-$n$ limits.

In section~\ref{sec:MHD-equilibrium}, we explain the ideal MHD equilibrium and stability equations. In section~\ref{sec:discretization}, we describe in detail how we discretize the ideal MHD energy principle into an eigenvalue problem. In section~\ref{sec:numerical-methods}, we explain the numerical methods that we use to solve the eigenvalue problem, including the algorithm and a discussion about when a computed eigenvalue is numerically relevant. In section~\ref{sec:benchmarks}, we benchmark our code against the~\texttt{NIMSTELL}~\cite{sovinec_semi-implicit_2026} solver, perform scans with increasing resolution, compare eigenvalue and gradient computation times across CPUs and GPUs, and present eigenvalue scans with the minor radius. In section~\ref{sec:incompressibility}, we present two different methods to impose incompressibility and show that they are numerically consistent with each other. In section~\ref{sec:conclusion}, we conclude our work and discuss the directions in which it can be extended.

\section{Ideal MHD equilibria and their linear stability}
\label{sec:MHD-equilibrium}
We begin with the visco-resistive single-fluid MHD model given by
\begin{equation}
    \frac{\partial n}{\partial t} + \bm{\nabla} \cdot (n \bm{V}) = 0,
    \label{eqn:MHD-continuity}
\end{equation}
\begin{equation}
    M n \left(\frac{\partial \bm{V}}{\partial t} + \bm{V}\cdot \bm{\nabla} \bm{V}\right) = -\bm{\nabla} p + \bm{j} \times \bm{B} - \bm{\nabla}\cdot \bm{\Pi},
    \label{eqn:MHD-momentum}
\end{equation}
\begin{equation}
    \mu_0 \bm{j} = \bm{\nabla} \times \bm{B} 
    \label{eqn:MHD-Ampere}
\end{equation}
\begin{equation}
    \frac{n}{\Gamma-1} \left( \frac{\partial T}{\partial t} + \bm{V}\cdot \bm{\nabla} T \right)  =  
    - n T \bm{\nabla}\cdot \bm{V} + \bm{\nabla}\cdot (\bm{\kappa} \cdot \bm{\nabla} T), 
    \label{eqn:MHD-temperature-evolution}
\end{equation}
\begin{equation}
    \frac{\partial \bm{B}}{\partial t} = \bm{\nabla} \times (\bm{V} \times \bm{B}) - \bm{\nabla} \times ( \eta \bm{j} ), 
    \label{eqn:MHD-induction}
\end{equation}
\begin{equation}
    \bm{\nabla}\cdot \bm{B} = 0,
    \label{eqn:div-B}
\end{equation}
where $n$ is the electron density, $\bm{V}$ is the mass-averaged flow velocity, $M$ is the effective mass per electron in a quasi-neutral plasma, $p$ is the total thermal pressure, $\bm{B}$ is the magnetic field, $\bm{j}$ is the current density, $\bm{\Pi}$ is the viscous stress tensor, $\mu_0$ is the vacuum magnetic permeability, $\Gamma$ is the ratio of specific heat capacities, $T$ is the plasma temperature assumed to be equal for electrons and ions (\emph{i.e.,} $p=(1+1/Z_i)nT$), $\bm{\kappa}$ is the thermal conductivity tensor, and $\eta$ is the electrical resistivity.

The MHD Ohm's law $\bm{E} = - \bm{V} \times \bm{B} + \eta \bm{j}$ has been used to obtain~\eqref{eqn:MHD-induction} from the induction equation.
We retain the full non-ideal system here to supplement the model description for the initial-value code~\texttt{NIMSTELL}, which is used for benchmarks in section~\ref{sec:benchmarks}. However, for an ideal MHD system, we have $\eta = 0, \bm{\Pi} = 0,$ and $\bm{\kappa} = 0$. 
The equations for a static equilibrium solver such as~\texttt{DESC} are obtained by assuming a steady state, $\partial /\partial t = 0$, and ignoring equilibrium flow, $\bm{V}_0 = 0$, so that
\begin{equation}
    \mathcal{F} \equiv - \bm{\nabla} p_0  + \bm{j}_0 \times \bm{B}_0 = 0, \
    \label{eqn:force-balance}
\end{equation}
\begin{equation}
    \bm{\nabla}\cdot \bm{B}_0 = 0,
    \label{eqn:div-B2}
\end{equation}
where the subscript 0 denotes equilibrium fields. The steady-state force balance equation~\eqref{eqn:force-balance} represents a balance between plasma pressure, magnetic pressure, and magnetic tension of magnetic field lines. The divergence-free condition~\eqref{eqn:div-B2} is an additional constraint that must be satisfied by the magnetic field vector throughout the plasma. For a general 3D geometry,~\eqref{eqn:div-B2} can be achieved by writing the magnetic field in the Clebsch form~\cite{d2012flux}.

After perturbing the fields and linearizing the ideal MHD model, the perturbed ideal MHD system can be written as
\begin{equation}
   -\bm{\nabla}\delta p + \delta\! \bm{j} \times \bm{B}_0 + \bm{j}_0  \times \delta\! \bm{B} =  M n_0 \frac{\partial \delta \bm{V}}{\partial t} , 
    \label{eqn:perturbed-Momentum}
\end{equation}
\begin{equation}
    \mu_0 \delta\! \bm{j} = \bm{\nabla} \times \delta\! \bm{B},
    \label{eqn:perturbed-Ampere}
\end{equation}
\begin{equation}
    \delta\! \bm{B} = \bm{\nabla} \times (\tilde{\bm{\xi}} \times \bm{B}_0),
    \label{eqn:perturbed-Induction}
\end{equation}
\begin{equation}
   \delta p = -(\Gamma p_0 \bm{\nabla}\cdot \tilde{\bm{\xi}} + \tilde{\bm{\xi}}\cdot \bm{\nabla} p_0)
    \label{eqn:perturbed-Energy}
\end{equation}
where $\delta$ denotes perturbed fields, $\delta\!p = \delta (2 n T)$ is the perturbed pressure, $\delta\! \bm{V} = \partial \tilde{\bm{\xi}}/\partial t$ is the perturbed velocity, and $\tilde{\bm{\xi}}$ is the physical displacement perturbation. Assuming a single eigenmode has the form $\tilde{\bm{\xi}}(\bm{r}) \exp(\gamma t)$ where $\gamma$ is the growth rate, and substituting this ansatz along with equations~\eqref{eqn:perturbed-Ampere}-\eqref{eqn:perturbed-Energy} into~\eqref{eqn:perturbed-Momentum}, the linearized momentum equation becomes a single eigenvalue equation for $\tilde{\bm{\xi}}$~\cite{bernsteinenergyprinciple}, and can be written as
\begin{equation}
     \bm{\mathrm{F}}[\tilde{\bm{\xi}}] = \gamma^2 M n_0 \tilde{\bm{\xi}}.
    \label{eqn:ideal-MHD-general-ODE}
\end{equation}
The left-hand-side in~\eqref{eqn:ideal-MHD-general-ODE} is known as the ideal MHD force operator
\begin{equation}
\begin{split}
    \bm{\mathrm{F}}[\tilde{\bm{\xi}}] &= \bm{\nabla}(\tilde{\bm{\xi}}\cdot\bm{\nabla}p_0 + \Gamma p_0 (\bm{\nabla} \cdot \tilde{\bm{\xi}})) + \frac{[\bm{\nabla} \times (\bm{\nabla}\times(\tilde{\bm{\xi}} \times \bm{B}_0))] \times \bm{B}_0}{\mu_0} \\
    &+ \frac{(\bm{\nabla} \times \bm{B}_0) \times (\bm{\nabla}\times(\tilde{\bm{\xi}} \times \bm{B}_0))}{\mu_0},
\end{split}
\end{equation}
whereas the right-hand-side denotes the acceleration of the plasma due to the perturbation $\tilde{\bm{\xi}}$. A detailed derivation of the perturbed ideal MHD model is given in Freidberg~\cite{IdealMHD}. By taking a dot product of $-\tilde{\bm{\xi}}$ with both sides of the linearized momentum equation~\eqref{eqn:ideal-MHD-general-ODE} and integrating throughout the volume, we can obtain the energy-principle formulation
\begin{equation}
   - \int dV \tilde{\bm{\xi}} \cdot \mathbf{F}[\tilde{\bm{\xi}}] = - \int dV\gamma^2 (M n_0) \lvert \tilde{\bm{\xi}} \rvert^2.
    \label{eqn:energy-integral}
\end{equation}
If we ignore vacuum (or external) modes, one can show that the left side of the energy integral~\eqref{eqn:energy-integral} can be written as
\begin{equation}
    \delta\! W_{\rm{p}} \equiv 
    \int dV \left(\bm{C}^2 + \Gamma p_0 \lvert \bm{\nabla}\cdot \tilde{\bm{\xi}} \rvert^2 - F \lvert \tilde{\bm{\xi}} \cdot \bm{\nabla} \rho \rvert^2  \right),
    \label{eqn:potential-energy}
\end{equation}
where
\begin{equation}
\begin{gathered}   
    \bm{C} = \bm{Q} + \frac{(\bm{j}_0 \times \bm{\nabla} \rho)}{\lvert \bm{\nabla} \rho \rvert^2}  \tilde{\bm{\xi}} \cdot \bm{\nabla}\rho,\\
    \bm{Q} = \bm{\nabla} \times (\tilde{\bm{\xi}} \times \bm{B}_0),\\
    F = \frac{2 (\bm{j}_0 \times \bm{\nabla}\rho) \cdot (\bm{B}_0\cdot \bm{\nabla} \nabla \rho)}{\lvert \bm{\nabla} \rho \rvert^4},
\end{gathered}
\end{equation}
and $\rho$ is a normalized radial coordinate. Note that all the gradients $\bm{\nabla}$ have been normalized using the average minor radius $a_{\rm{N}}$ and the magnetic field $\bm{B}_0$ using $B_{\rm{N}}$. The constant $B_{\rm{N}}$ has been chosen so that $B_{\rm{N}} = |\psi_{\rm{b}}|/(\pi{a_N}^2)$ where $\psi_{\rm{b}}$ is the toroidal magnetic flux enclosed by the boundary. The potential energy integral is taken from Bernstein~\textit{et al.}~\cite{bernsteinenergyprinciple} but follows the notation in the~\texttt{TERPSICHORE}~\cite{Anderson1990TERPSICHORE} code. 
The normalized kinetic energy integral
\begin{equation}
    \delta K = \int dV \lvert \tilde{\bm{\xi}} \rvert^2,
\end{equation}
is positive definite. The problem is to solve for a mode $\tilde{\bm{\xi}}$ and an eigenvalue $\lambda$, such that
\begin{equation}
    \delta W_{\mathrm{p}} = -\lambda\, \delta\! K,\\
    \label{eqn:stability-problem-1}
\end{equation}
where the eigenvalue $\lambda = \gamma^2 (a_{\rm{N}}/v_{A})^2$ and $v_{A} = B_{\rm{N}}/\sqrt{\mu_0 M n_0}$ is the normalizing Alfv\'en speed. The eigenvalue informs us about the stability of an equilibrium, 
$\lambda > 0$ implies instability as two real values of $\gamma$ are possible, one of which must be positive, whereas $\lambda < 0$ corresponds to a stable equilibrium with two imaginary values of $\gamma$.\footnote{The normalization and sign convention is chosen to ensure consistency with the infinite-$n$ ideal ballooning solver~\cite{Gaur2025OmnigenousStability} in~\texttt{DESC}.}
In the next section, we provide the numerical scheme for discretizing and solving this problem.

\section{Discretization using differentiation matrices}
\label{sec:discretization}
We focus on solutions whose magnetic field lines lie on closed nested toroidal surfaces, known as flux surfaces. We label these surfaces using the enclosed toroidal flux $\psi$. Each flux surface is labeled by the normalized toroidal flux $\rho = \sqrt{\psi/\psi_{\mathrm{b}}}$, where $\psi_{\mathrm{b}}$ is the toroidal flux enclosed by the boundary so that $\rho \in [0, 1]$.
The form of a divergence-free field in a nested toroidal coordinate system $(\rho, \theta, \zeta)$ is  
\begin{equation}
    \bm{B}_0 = \frac{\psi^{'}}{\sqrt{g}} (\iota \bm{e}_{\theta} + \bm{e}_{\zeta}), 
\end{equation}
where $\psi^{'} = d\psi/d\rho$, $\theta$ is the PEST~\cite{Grimm1976PEST} angle and $\zeta$ is the cylindrical toroidal angle\footnote{The formalism presented in this paper can be easily generalized to any straight field line coordinate system. We choose the PEST coordinate system because of its robustness and computational speed.}, $\sqrt{g} = (\bm{\nabla}\rho \times \bm{\nabla}\theta \cdot \bm{\nabla}\zeta)^{-1}$ is the Jacobian of the curvilinear toroidal coordinate system, and $\iota$ is the rotational transform, the average pitch of a field line on a flux surface. The equilibrium current is
\begin{equation}
    \bm{j}_0 = j^{\theta} \bm{e}_{\theta} + j^{\zeta} \bm{e}_{\zeta}.
\end{equation}
We will write the vector $\tilde{\bm{\xi}}$ as
\begin{equation}
    \tilde{\bm{\xi}} = \tilde{\xi^{\rho}} \bm{e}_{\rho} + \tilde{\xi^{\theta}} \bm{e}_{\theta} + \tilde{\xi^{\zeta}} \bm{e}_{\zeta},
\end{equation}
so that
\begin{equation}
\begin{gathered}
    \tilde{\bm{\xi}} \times \bm{B}_0 = \psi^{'}\iota \tilde{\xi^{\rho}} \frac{\bm{e}_{\rho} \times \bm{e}_{\theta}}{\sqrt{g}} -  \psi^{'}\tilde{\xi^{\rho}} \frac{\bm{e}_{\zeta} \times \bm{e}_{\rho}}{\sqrt{g}}  + \psi^{'}\tilde{\xi^{\theta}} \frac{\bm{e}_{\theta} \times \bm{e}_{\zeta}}{\sqrt{g}} - \iota \psi^{'} \tilde{\xi^{\zeta}} \frac{\bm{e}_{\theta} \times \bm{e}_{\zeta}}{\sqrt{g}}\\
    = \psi^{'}(\tilde{\xi^{\theta}} - \iota \tilde{\xi^{\zeta}}) \bm{e}^{\rho} - \psi^{'} \tilde{\xi^{\rho}} \bm{e}^{\theta} + \iota \psi^{'} \tilde{\xi^{\rho}} \bm{e}^{\zeta}.
\end{gathered}
\label{eqn:xi_cross_B}
\end{equation}
Before proceeding further, we note that some of the terms in quantities such as $\bm{Q}^2$ can behave poorly near the magnetic axis. To regularize these terms, we rescale the perturbation as
\begin{equation}
\begin{split}
    \xi^{\rho} &=  \frac{\tilde{\xi^{\rho}}}{\psi^{'}},\\
    \xi^{\theta} &=  \tilde{\xi^{\theta}}, \\
    \xi^{\zeta} &= \iota \tilde{\xi^{\zeta}}.
    \label{eqn:scaled-xi}
\end{split}
\end{equation}
This rescaling serves two purposes. First, since $\psi^{'} \propto \rho$ near the magnetic axis, it ensures $\lim_{\rho \rightarrow 0} \tilde{\xi^{\rho}} = 0$, assuming $\xi^{\rho}$ is finite. Second, in the $\bm{e}_{\zeta}$ component of $\bm{Q}$ in~\eqref{eqn:Q1}, every occurrence of $\tilde{\xi^{\rho}}$ is replaced by $\psi^{'} \xi^{\rho}$, which compensates the $1/\rho$ behavior associated with the $1/\sqrt{g}$ factor and keeps the corresponding terms finite and regular near the magnetic axis. Substituting the scaled $\bm{\xi}$ from~\eqref{eqn:scaled-xi} into~\eqref{eqn:xi_cross_B}, the field line bending term $\bm{Q} = \bm{\nabla} \times (\bm{\xi} \times \bm{B}_0)$ can be calculated as
\begin{equation}
    \bm{Q} = \frac{{\psi^{'}}^2}{\sqrt{g}}(\iota \partial_{\theta} + \partial_{\zeta}) \xi^{\rho} \bm{e}_{\rho} + \frac{\psi^{'}}{\sqrt{g}} \left( \partial_{\zeta}\upsilon - \frac{\partial_{\rho}(\iota {\psi^{'}}^2\xi^{\rho})}{\psi^{'}}\right) \bm{e}_{\theta} - \frac{\psi^{'}}{\sqrt{g}}\left(\partial_{\theta}\upsilon + \frac{\partial_{\rho} ({\psi^{'}}^2 \xi^{\rho})}{\psi^{'}}\right) \bm{e}_{\zeta},
    \label{eqn:Q1}
\end{equation}
with $\upsilon = \xi^{\theta} - \xi^{\zeta}$. It is numerically convenient to formulate the problem in $(\xi^\rho, \upsilon, \xi^\zeta)$ coordinates but transforming from $\upsilon$ to $\xi^{\theta}$ is straightforward. In the next section, we explain how the derivative terms, such as the one in~\eqref{eqn:Q1} are calculated numerically using differentiation matrix operators.


\subsection{Differentiation matrices and quadratures}
We now discuss the scheme used to discretize the energy integral. Every term in the general form of the energy integral can be expressed as a quadrature in the following way
\begin{equation}
    \int_{0}^{1} \int_{0}^{2\pi} \int_{0}^{2\pi} d\rho d\theta d\zeta \, \sqrt{g}\, \partial_{y} \xi^{x} \, \mathrm{E} \, \partial_{{y^{'}}} \xi^{x^{'}} = \sum_{\rho} w_{\rho} \sum_{\theta} w_{\theta} \sum_{\zeta} w_{\zeta}  \mathrm{E} \sqrt{g} \, \partial_{y} \xi^{x}(\rho_i, \theta_i, \zeta_i) \, \partial_{y^{'}} \xi^{x^{'}}(\rho_i, \theta_i, \zeta_i),
    \label{eqn:integral-quadrature}
\end{equation}
where the weights ($w_{\rho}, w_{\theta}, w_{\zeta}$) and the ordered grid of collocation nodes $(\rho_i, \theta_i, \zeta_i)$ depend on the type of quadrature used.
We utilize the differentiation matrix method~\cite{trefethen2000spectral, canuto2006spectral} and write a derivative $\partial_{y} \xi^{x} \approx D_{y} \xi^{x}$,  
where $x, x', y, y' \in (\rho, \theta, \zeta)$ to indicate the component of $\bm{\xi}$ and the direction in which the derivative is taken. The matrix $D$ is a first-order spectral differentiation matrix and $\rm{E}$  and $\sqrt{g}$ are purely equilibrium-dependent quantities evaluated at the collocation points. Each term can be written in this form, and the symmetry of the energy integral guarantees that for each off-diagonal term in the integrand there exists a conjugate term so that the discretized matrix is always Hermitian. 

Since the product of matrices corresponding to different dimensions is a Kronecker product, the final matrix will be of the shape $3N \times 3N$ where $N = \nr \nt \nz$ is the total number of collocation points, $\nr, \nt$, and $\nz$, being the radial, poloidal, and toroidal resolution values, respectively. There are nine major blocks corresponding to different combinations of the components of the perturbation $\bm{\xi}$. In each block, there are nine smaller blocks corresponding to derivatives with respect to the coordinates. For example, after applying the differentiation matrix discretization method to~\eqref{eqn:integral-quadrature} and choosing $x = y^{'} = \zeta, x^{'} = y = \rho$ we get
\begin{equation}
    \partial_{\rho} \xi^{\zeta} (W \sqrt{g} E) \partial_{\zeta} \xi^{\rho} \approx {\xi^{\zeta}}^{\dagger}(D_{\rho,0}^{\dagger} \otimes \mathbb{I}_{\theta,0} \otimes \mathbb{I}_{\zeta,0}) \mathrm{diag}(W  \sqrt{g}\, \mathrm{E})\,  (\mathbb{I}_{\rho,0} \otimes \mathbb{I}_{\theta,0} \otimes D_{\zeta,0})\, {\xi^{\rho}}.
\end{equation} 
This term will live in the $\zeta-\rho$ block which is determined by the displacement components on left and right side, $\xi^{\zeta}$ and $\xi^{\rho}$, respectively, and has size $N \times N$. For notational simplicity, we define 
\begin{equation}
\begin{split}
    D_{\rho}   &\equiv  D_{\rho,0} \otimes \mathbb{I}_{\theta,0} \otimes \mathbb{I}_{\zeta,0},\\
    D_{\theta} &\equiv  \mathbb{I}_{\rho,0} \otimes D_{\theta,0} \otimes \mathbb{I}_{\zeta,0}, \\
    D_{\zeta}  &\equiv  \mathbb{I}_{\rho,0} \otimes \mathbb{I}_{\theta,0} \otimes D_{\zeta,0},
    \label{eqn:diffmatrices}
\end{split}
\end{equation}
where the subscript $0$ is used to denote the smaller matrices of size $\nr \times \nr$, $\nt \times \nt$, or $\nz \times \nz$. Note that for tokamaks, we use $D_{\zeta,0} = i n$ and for axisymmetric mirrors, we use $D_{\theta,0} = i m$ where $m$ and $n$ are the poloidal and toroidal mode numbers, respectively, which are specified by the user. The axisymmetry (homogeneity) makes the differentiation matrices diagonal with all entries being the same. For these cases, each mode number can be solved independently of other toroidal mode numbers which makes the final matrix much smaller, by a factor of $\nt$ for an axisymmetric mirror and $\nz$ for a tokamak, and the stability calculation significantly faster.

\subsection{Bases and mapping functions}
The natural basis used to create differentiation matrices and quadratures in the toroidal and poloidal directions is the Fourier basis, due to the double periodicity of a toroidal geometry. For the radial direction, the default option is a Gauss-Jacobi-Radau (GJR) basis unless specified otherwise. However, we offer a variety of different bases including Legendre-Lobatto, B-spline, finite-difference, and Zernike. Further information about the different quadrature schemes and related tests of the derivatives are provided in appendix~\ref{app:Bases}.

The GJR points span $[-1, 1)$ and concentrate near the ends. We perform all the calculations on a transformed grid $\rho_{\mathrm{s}}$ so that $\rho_{\rm{s}} = f(\rho)$ such that $\rho_{\rm{s}} \in [0, 1]$ and the mapping function $f$  allows collocation points to move towards some target radius $x_{0}$ of interest (more information in appendix~\ref{subsec:radial-mapping-function}). In other words, the quadrature nodes and weights live on the GJR grid $\rho$ and the integrand and equilibrium quantities are evaluated at the mapped points $\rho_{\rm{s}} = f(\rho)$. After remapping $\rho$, the integral on the modified grid $\rho_{\rm{s}}$ can be written as an integral on the original radial grid $\rho$ as
\begin{equation}
    \int_{0}^{1} d\rho_{\rm{s}}  \, X(\rho_{\rm{s}}) = \int_{-1}^{1}  d\rho f^{'}(\rho) X(f(\rho)).
\end{equation}
This will transform the quadrature points and weights as well as the radial differentiation matrix, and evaluate the equilibrium quantities at the points $\rho_{\rm{s}}$. The quadrature corresponding to a general term of the energy integral, after discretization will then give us
\begin{equation}
    \int \sqrt{g}\, d\rho_{\rm{s}}\, d\theta\, d\zeta\, E\, \partial_{y} \xi^{x}\, \partial_{y'} \xi^{x'} \approx {\xi^{x}}^{\dagger} [{D_{y}}^{\dagger} \rm{diag}(W_{\rm{s}} W  \sqrt{g}\, \mathrm{E}) D_{y^{'}}] {\xi^{x}}^{'}, y, y^{'} \neq \rho,
    \label{eqn:discretization}
\end{equation} 
where all the quantities are evaluated at the collocation nodes $(\rho_{\rm{s}i}, \theta_i, \zeta_i)$ and $W_{\rm{s}} = \operatorname{diag}(f'(\rho)) \otimes \mathbb{I}_{\theta, 0} \otimes \mathbb{I}_{\zeta, 0}$ is a diagonal matrix. The exact expression of the mapping function is provided in appendix~\ref{subsec:radial-mapping-function}.

\subsection{Expanding the energy integral formulation}
We can now expand each term in the energy integral~\eqref{eqn:potential-energy}. In the new curvilinear coordinate system $(\rho_{\rm{s}}, \theta, \zeta)$ we first expand the term
\begin{equation}
    \bm{C}^2 = \bm{Q}^2 + \left(\frac{\bm{j}_0 \times \bm{\nabla}\rho}{\lvert \bm{\nabla}\rho \rvert^2} \right)^2 {\xi^{\rho}}^2 +  {\xi^{\rho}}^{T}\left(\frac{\bm{j}_0 \times \bm{\nabla}\rho}{\lvert \bm{\nabla}\rho \rvert^2}\right)  \cdot \bm{Q} +   \bm{Q} \cdot \left(\frac{\bm{j}_0 \times \bm{\nabla}\rho}{\lvert \bm{\nabla}\rho \rvert^2}\right) \xi^{\rho}.
    \label{eqn:C1}
\end{equation}
With the quadrature in curvilinear coordinates explained, we discretize the different terms in the square of the field line bending term $\bm{Q}^2$. We define $(\bm{Q}^2)_{xx'}$ as the dot product of the $x$ and $x'$ contravariant components of $\bm{Q}$ in~\eqref{eqn:Q1}. Using this convention we can write various components as  
\begin{equation}
    (\bm{Q}^2)_{\rho \rho} \equiv g_{\rho \rho} \left(\frac{{\psi^{'}}^2}{\sqrt{g}}\right)^2 \left[\iota^2 (\partial_{\theta} \xi^{\rho_{\rm{s}}})^2 + (\partial_{\zeta} \xi^{\rho_{\rm{s}}})^2 + 2 \iota (\partial_{\theta} \xi^{\rho_{\rm{s}}})(\partial_{\zeta} \xi^{\rho_{\rm{s}}}) \right],
    \label{eqn:Q2_rho-rho}
\end{equation}
\begin{equation}
\begin{split}
    (\bm{Q}^2)_{\theta \theta} \equiv g_{\theta \theta} \left(\frac{{\psi^{'}}}{\sqrt{g}}\right)^2 \Bigg[&(\partial_{\zeta} \upsilon)^2 + \left(\frac{\partial_{\rho_{\rm{s}}}(\iota {\psi^{'}}^2 \xi^{\rho_{\rm{s}}})}{\psi^{'}}\right)^2 -2 (\partial_{\zeta}\upsilon) \frac{\partial_{\rho_{\rm{s}}}(\iota {\psi^{'}}^2 \xi^{\rho_{\rm{s}}})}{\psi^{'}} \Bigg],
    \label{eqn:Q2_theta-theta}
\end{split}
\end{equation}

\begin{equation}
\begin{split}
    (\bm{Q}^2)_{\zeta \zeta} \equiv g_{\zeta \zeta} \left(\frac{{\psi^{'}}}{\sqrt{g}}\right)^2 \Bigg[&(\partial_{\theta} \upsilon)^2  + \left(\frac{\partial_{\rho_{\rm{s}}}({\psi^{'}}^2 \xi^{\rho_{\rm{s}}})}{\psi^{'}}\right)^2  +2 (\partial_{\theta}\upsilon) \frac{\partial_{\rho_{\rm{s}}}({\psi^{'}}^2 \xi^{\rho_{\rm{s}}})}{\psi^{'}} \Bigg],
    \label{eqn:Q2_zeta-zeta}
\end{split}
\end{equation}

\begin{equation}
\begin{split}
    (\bm{Q}^2)_{\rho \theta} \equiv  g_{\rho \theta} \frac{{\psi^{'}}^3}{\sqrt{g}^2} \Bigg[& \iota (\partial_{\theta} \xi^{\rho_{\rm{s}}}) (\partial_{\zeta} \upsilon)  + (\partial_{\zeta} \xi^{\rho_{\rm{s}}}) (\partial_{\zeta} \upsilon) \\
    &-  \iota (\partial_{\theta} \xi^{\rho_{\rm{s}}}) \left(\frac{\partial_{\rho_{\rm{s}}}(\iota {\psi^{'}}^2 \xi^{\rho_{\rm{s}}})}{\psi^{'}}\right) - (\partial_{\zeta} \xi^{\rho_{\rm{s}}})  \left(\frac{\partial_{\rho_{\rm{s}}}(\iota {\psi^{'}}^2 \xi^{\rho_{\rm{s}}})}{\psi^{'}}\right)\Bigg],
    \label{eqn:Q2_rho-theta}
\end{split}
\end{equation}

\begin{equation}
\begin{split}
    (\bm{Q}^2)_{\rho \zeta} \equiv -  g_{\rho \zeta} \frac{{\psi^{'}}^3}{\sqrt{g}^2} \Bigg[& \iota (\partial_{\theta} \xi^{\rho_{\rm{s}}}) (\partial_{\theta} \upsilon)  + (\partial_{\zeta} \xi^{\rho_{\rm{s}}}) (\partial_{\theta} \upsilon) \\
    &+  \iota (\partial_{\theta} \xi^{\rho_{\rm{s}}}) \left(\frac{\partial_{\rho_{\rm{s}}}({\psi^{'}}^2 \xi^{\rho_{\rm{s}}})}{\psi^{'}}\right) + (\partial_{\zeta} \xi^{\rho_{\rm{s}}})  \left(\frac{\partial_{\rho_{\rm{s}}}({\psi^{'}}^2 \xi^{\rho_{\rm{s}}})}{\psi^{'}}\right)\Bigg],
    \label{eqn:Q2_rho-zeta}
\end{split}
\end{equation}

\begin{equation}
\begin{split}
    (\bm{Q}^2)_{\theta \zeta} \equiv - g_{\theta \zeta} \left(\frac{{\psi^{'}}}{\sqrt{g}}\right)^2 \Bigg[& (\partial_{\zeta} \upsilon) (\partial_{\theta} \upsilon) + (\partial_{\zeta} \upsilon) \left(\frac{\partial_{\rho_{\rm{s}}}({\psi^{'}}^2 \xi^{\rho_{\rm{s}}})}{\psi^{'}}\right) - (\partial_{\theta} \upsilon) \left(\frac{\partial_{\rho_{\rm{s}}}(\iota {\psi^{'}}^2 \xi^{\rho_{\rm{s}}})}{\psi^{'}}\right)\\ 
    &-\left(\frac{\partial_{\rho_{\rm{s}}}(\iota {\psi^{'}}^2 \xi^{\rho_{\rm{s}}})}{\psi^{'}}\right)   \left(\frac{\partial_{\rho_{\rm{s}}}({\psi^{'}}^2 \xi^{\rho_{\rm{s}}})}{\psi^{'}}\right) \Bigg].
    \label{eqn:Q2_theta-zeta}
\end{split}
\end{equation}
The field line bending term $\bm{Q}^2 = \sum_{x, x=x'}(\bm{Q}^2)_{x x'} + \sum_{x, x < x', x\neq x'} ((\bm{Q}^2)_{x x'} + [(\bm{Q}^2)_{x x'}]^{\dagger})$ where $x, x' \in \{\rho, \theta, \zeta\}$.  After $\bm{Q}^2$, we evaluate the next squared term in $\bm{C}^2$,
\begin{equation}
\int dV \left(\frac{\bm{j}_0 \times \bm{\nabla}\rho}{\lvert \bm{\nabla}\rho \rvert^2}\right)^2 (\bm{\xi}\cdot \bm{\nabla}\rho)^2 \approx {\xi^{\rho_{\rm{s}}}}^{\dagger}\left( \frac{{\psi^{'}}^2 W \sqrt{g} |\bm{j}|^2}{g^{\rho \rho}} \right){\xi^{\rho_{\rm{s}}}},
\end{equation}
where we assumed $\bm{j}_0\cdot\bm{\nabla}\rho = 0$ as a consequence of a nested ideal MHD equilibrium assumption. Before writing the mixed terms in $\bm{C}^2$, we use the expansion 
\begin{equation}
\begin{split}
    (\bm{j}_0 \times \bm{\nabla}\rho)/|\bm{\nabla}\rho|^2 = \sqrt{g} \left[ (j^{\theta} g^{\zeta \rho} - j^{\zeta} g^{\theta \rho})\frac{\bm{\nabla}\rho}{|\bm{\nabla}\rho|^2} + j^{\zeta} \bm{\nabla}\theta - j^{\theta} \bm{\nabla}\zeta \right].
\end{split}
\end{equation}
to obtain the mixed term 
\begin{equation}
\begin{split}
  \int dV (\tilde{\bm{\xi}}\cdot \bm{\nabla}\rho)\left(\frac{\bm{j}_0 \times \bm{\nabla}\rho}{\lvert \bm{\nabla}\rho \rvert^2}\right)  \cdot \bm{Q} &=  {\xi^{\rho_{\rm{s}}}} \left[  {\psi^{'}}^{3} W \sqrt{g} \frac{(j^{\theta} g^{\zeta \rho} - j^{\zeta} g^{\theta \rho})}{g^{\rho\rho}} (\iota \partial_{\theta} + \partial_{\zeta}) \right] {\xi^{\rho_{\rm{s}}}} \\
  &+ {\xi^{\rho_{\rm{s}}}}  {\psi^{'}}^2 W \sqrt{g} \left(j^{\zeta} \partial_{\zeta} + j^{\theta} \partial_{\theta} \right)\upsilon\\
  &- {\xi^{\rho_{\rm{s}}}}  W {\psi^{'}} \sqrt{g} \left[ j^{\zeta} \partial_{\rho_{\rm{s}}}(\iota {\psi^{'}}^2 \xi^{\rho_{\rm{s}}}) - j^{\theta}\partial_{\rho_{\rm{s}}}({\psi^{'}}^2 \xi^{\rho_{\rm{s}}})\right]
  \label{eqn:mixed}
\end{split}
\end{equation}
and similarly calculate its complex conjugate (not shown). Note that we utilize the force balance relation to obtain $j^{\theta} = \iota j^{\zeta} + p_0^{'}/\psi^{'}$ and the identities that can give us the contravariant metric elements $g^{\zeta \rho}, g^{\theta \rho}$ in terms of the covariant elements and the Jacobian. This reduces the number of inputs to the solver. Next, we expand the compressibility term 
\begin{equation}
    \Gamma p_0 (\bm{\nabla} \cdot \tilde{\bm{\xi}})^2 = \Gamma p_0 \left\{\frac{1}{\sqrt{g}}\left[\frac{\partial (\sqrt{g} \psi^{'} \xi^{\rho_{\rm{s}}})}{\partial \rho_{\rm{s}}} + \frac{\partial (\sqrt{g} (\upsilon+\xi^{\zeta}))}{\partial \theta} + \frac{\partial (\sqrt{g} \xi^{\zeta}/\iota)}{\partial \zeta} \right]\right\}^2,
    \label{eqn:div-xi-term}
\end{equation}
where we use~\eqref{eqn:scaled-xi} in~\eqref{eqn:div-xi-term}. Finally, we discretize the instability drive term
\begin{equation}
    \int dV F (\tilde{\bm{\xi}}\cdot \bm{\nabla}\rho)^2 \approx \xi^{\rho_s \dagger} (W \sqrt{g} F {\psi^{'}}^2) \xi^{\rho_s}.
\end{equation}
For the ease of readability, we have moved the discretized expressions corresponding to the fieldline bending terms~\eqref{eqn:Q2_rho-rho}-\eqref{eqn:Q2_theta-zeta}, the mixed term~\eqref{eqn:mixed}, and the compressibility term~\eqref{eqn:div-xi-term} to appendix~\ref{app:Q2}.  
For the kinetic energy integral, the discretized form is
\begin{equation}
\begin{split}
    \int dV \, \lvert\tilde{\bm{\xi}}\rvert^2 &\approx \xi^{\rho_s \dagger} (W \sqrt{g}\, g_{\rho \rho} {\psi^{'}}^2) \xi^{\rho_s} + (\upsilon + \xi^{\zeta})^{\dagger} (W \sqrt{g}\, g_{\theta \theta}) (\upsilon + \xi^{\zeta}) \\
    &+ \xi^{\zeta \dagger} ( W \sqrt{g}\, g_{\zeta \zeta}/\iota^2) \xi^{\zeta}+ \xi^{\rho_s \dagger} ( W \sqrt{g}\, g_{\rho \theta} \psi^{'}) (\upsilon + \xi^{\zeta})\\
    &+ \xi^{\rho_s \dagger} (W \sqrt{g}\, g_{\rho \zeta} \psi^{'}/\iota) \xi^{\zeta} +  (\upsilon + \xi^{\zeta})^{\dagger}  (W \sqrt{g}\, g_{\theta \zeta}/\iota) \xi^{\zeta} + \mathrm{c.c}.
\end{split}
\end{equation}
where $\mathrm{c.c.}$ is the complex conjugate of all the off-diagonal terms. After all the terms on both sides of~\eqref{eqn:stability-problem-1} are summed 
and the right side negated (to account for the ``$-$'' sign in front of $\lambda$ in~\eqref{eqn:stability-problem-1}), we form a discretized variational principle
\begin{equation}
    \langle \bm{\xi}| \mathbb{A} | \bm{\xi} \rangle = \lambda \langle \bm{\xi}| \mathbb{B} | \bm{\xi} \rangle,
\end{equation}
where $\mathbb{A}$ is the potential energy matrix and $\mathbb{B}$ is the kinetic energy matrix and the inner product $\langle x\lvert y \rangle = x^{\dagger} y$ is defined in a bounded basis with a well-defined norm.
When expanded, the term on the left side of the equation can be written as
\[
\bm{\xi}^{\dagger} \mathbb{A} \bm{\xi} = 
[\bar{\xi^{\rho_s}}, \bar{\upsilon}, \bar{\xi^{\zeta}}]
\begin{pmatrix} 
A_{\rho_s \rho_s}   & A_{\rho_s \theta}   & A_{\rho_s \zeta} \\ 
A_{\theta \rho_s} & A_{\theta \theta} & A_{\theta \zeta} \\ 
A_{\zeta \rho_s}  & A_{\zeta \theta}  & A_{\zeta \zeta} 
\end{pmatrix}
\left[
\begin{matrix} 
\xi^{\rho_s}\\
\upsilon\\
\xi^{\zeta}
\end{matrix}
\right],
\]
and the term on the right side of the equation becomes
\[
\bm{\xi}^{\dagger} \rm{diag}(\lambda \mathbb{I})\mathbb{B} \bm{\xi} = 
[\bar{\xi^{\rho_s}}, \bar{\upsilon}, \bar{\xi^{\zeta}}]
\begin{pmatrix} 
\lambda \mathbb{I} & 0        & 0 \\ 
0       & \lambda  \mathbb{I} & 0 \\ 
0       &    0     & \lambda \mathbb{I}
\end{pmatrix}
\begin{pmatrix} 
B_{\rho_s \rho_s}   & B_{\rho_s \theta}   & B_{\rho_s \zeta} \\ 
B_{\theta \rho_s} & B_{\theta \theta} & B_{\theta \zeta} \\ 
B_{\zeta \rho_s}  & B_{\zeta \theta}  & B_{\zeta \zeta} 
\end{pmatrix}
\left[
\begin{matrix} 
\xi^{\rho_s}\\
\upsilon\\
\xi^{\zeta}
\end{matrix}
\right],
\]
where $\bar{x}$ is the complex conjugate of $x$. Each block $A_{xx'}$, $B_{xx'}\, \forall \, x, x' \in \{\rho_s, \theta, \zeta\}$, and the identity matrix are matrices of size $N = N_{\rho} N_{\theta} N_{\zeta}$ and due to the Hermitian structure of the problem, $A_{xx'} = {A}^{\dagger}_{x'x}, B_{xx'} = {B}^{\dagger}_{x'x}$. Since the matrix $\mathbb{B}$ corresponds to the integral of the kinetic energy, it is Hermitian positive definite with all the blocks ${B}^{\dagger}_{xx'}$ being purely diagonal. Since the matrices are Hermitian, solutions of the variational principle will also be the solutions to the generalized Hermitian eigenvalue problem
\begin{equation}
     \mathbb{A} \bm{\xi} = \lambda  \mathbb{B}  \bm{\xi}.
\end{equation}

The input equilibrium quantities supplied to~\texttt{AGNI} and the resulting output quantities after solving the eigenvalue problem are summarized in Table~\ref{tab:solver_interface}. Note that the code will work for any coordinate system as long as the input quantities are calculated in a straight field line coordinate system.
\begin{table}[H]
\centering
\caption{Inputs and outputs of \texttt{AGNI}. The index
$i=1,\ldots,k$ is the rank $i$ unstable eigenmode.}
\label{tab:solver_interface}
\begin{tabular}{lp{0.55\textwidth}l}
\hline
 & Quantity & Dependence \\
\hline
Inputs
& $a_{\rm{N}}, B_{\rm{N}}$
& none (scalars) \\
 
& $\iota$, $\psi'$, $p_0$, $p^{'}_0$, $n_0$ 
& $\rho$ \\

& $\sqrt{g}$, $\partial_{\rho}\sqrt{g}$, $\partial_{\theta}\sqrt{g}$, $\partial_{\zeta}\sqrt{g}$, $g_{\rho\rho}$, $g_{\theta\theta}$,
  $g_{\zeta\zeta}$, $g_{\rho\theta}$, $g_{\rho\zeta}$,
  $g_{\theta\zeta}$, $j^\zeta$, $F$
& $(\rho,\theta,\zeta)$ \\
\hline
Outputs
& $\lambda_i$
& none (scalar) \\

& $\xi_i^\rho$, $\xi_i^\theta$, $\xi_i^\zeta$,
  $\lvert\bm{Q}_i\rvert$,
  $|\delta\mathbf{V}_i|$
& $(\rho,\theta,\zeta)$ \\
\hline
\end{tabular}
\end{table}
The details about simplifying the generalized eigenvalue problem to a standard eigenvalue problem, and the rest of the numerical methods are explained in the following section.

\section{Numerical methods}
\label{sec:numerical-methods}
In this section, we will describe how we can use the structure of the kinetic energy matrix to convert the generalized Hermitian eigenvalue problem into a standard Hermitian eigenvalue problem. Next, we will explain the boundary conditions and how they are imposed. We will then look at the full eigenvalue spectrum of a stellarator and discuss the accuracy to which we can resolve the eigenvalue of an equilibrium. We will follow that by explaining how we use the shift-invert and Jacobi--Davidson methods to efficiently find the most unstable mode which makes the calculation significantly faster. Finally, we explain the matrix-free method and how it makes the solver more memory efficient.

\subsection{Efficient Cholesky decomposition of the kinetic energy matrix}
\label{subsec:efficient-Cholesky}
\begin{figure}
    \centering
    \begin{subfigure}[b]{0.32\textwidth}
    \centering
        \includegraphics[width=\textwidth, trim={2mm 2mm 1mm 2mm}, clip]{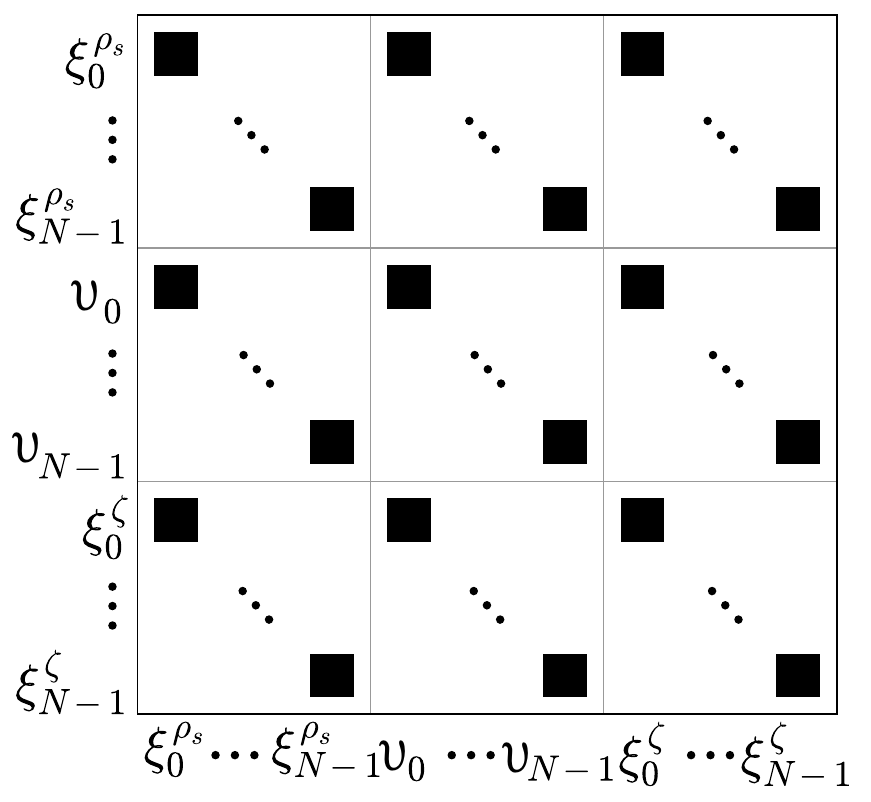}
        \caption{Component-major ordering}
    \end{subfigure}
    \qquad \qquad \qquad
    \begin{subfigure}[b]{0.32\textwidth}
        \centering
        \includegraphics[width=\textwidth, trim={2mm 2mm 2mm 2mm}, clip]{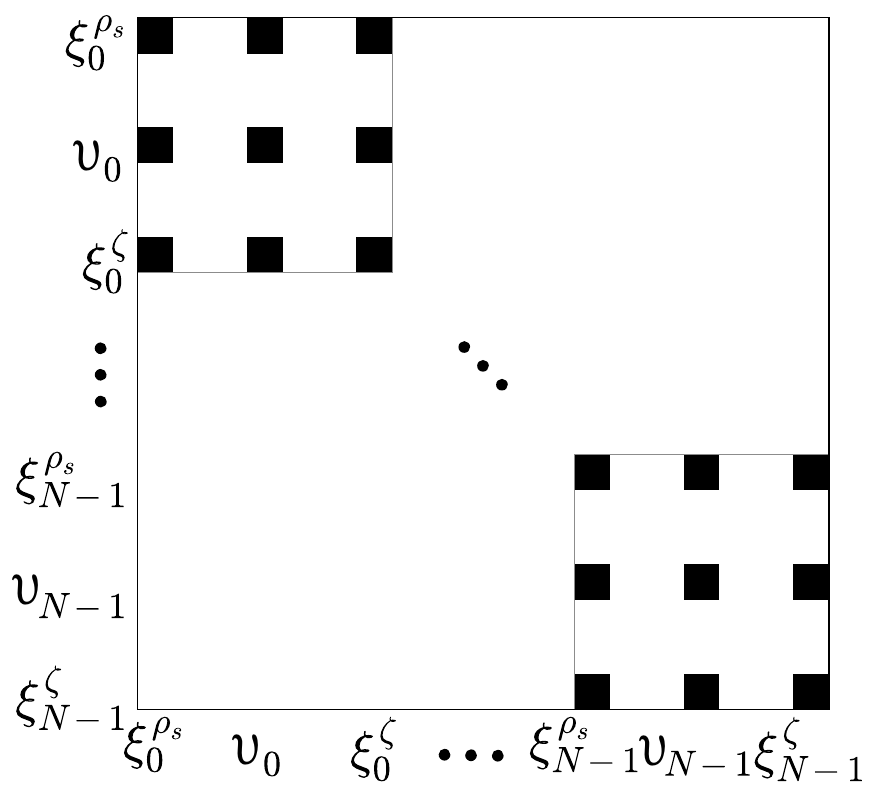}
        \caption{Node-major ordering}
        \label{fig:Node-major}
    \end{subfigure}
\caption{We present two equivalent ways of representing the kinetic energy matrix $\mathbb{B}$. Each black square is a nonzero entry. In figure~\textit{(a)}, we arrange the matrix by each component (of $\bm{\xi}$) value at all nodes whereas in figure~\textit{(b)} we arrange it by each collocation node with all its components. Both forms are equivalent but this rearrangement allows us to go from~\textit{(a)} where all blocks are diagonal to~\textit{(b)}, which is purely block diagonal. Hence, decomposing~\textit{(b)} is equivalent to decomposing $N$ $3 \times 3$ independent matrices. This reduces the computational complexity of Cholesky decomposition of $\mathbb{B}$ from $\mathcal{O}(N^3)$ to $\mathcal{O}(N)$.}
\label{fig:Component-to-row-major-ordering}
\end{figure}

The symmetric positive definite kinetic energy matrix comprises nine diagonal blocks $B_{xx'}\, \forall \, x, x' \in \{\rho_{\rm{s}}, \theta, \zeta\}$, each with a shape $N \times N$. Physically, the diagonal form is a consequence of the fact that without any derivatives, the components of the vector $\bm{\xi}$ at each node are decoupled from the rest of the nodes. This allows us to invert or decompose the inertia matrix $\mathbb{B}$ easily. The efficient way to Cholesky decompose the matrix $\mathbb{B}$ is illustrated in figure~\ref{fig:Component-to-row-major-ordering}.

Cholesky decomposition of $\mathbb{B} = \mathbb{L} \mathbb{L}^T$ allows us to transform from the physical coordinate $\bm{\xi}$ to computation coordinate $\bm{v} = \mathbb{L}^{T} \bm{\xi}$ which modifies the generalized Hermitian eigenvalue problem to a standard Hermitian eigenvalue problem
\begin{equation}
    \hat{\mathbb{A}}  \bm{v} =  \lambda \bm{v}, \quad \hat{\mathbb{A}} = \mathbb{L}^{-1} \mathbb{A} \mathbb{L}^{-T}.
    \label{eqn:standard-EVP}
\end{equation}
In the next section, we explain how to apply boundary conditions to the stability eigenvalue problem.

\subsection{Boundary conditions}
The stability problem is solved subject to Dirichlet boundary conditions (perfectly conducting wall) on the perturbed radial displacement $\xi^{\rho_{\rm{s}}}(\rho_{\rm{s}} = 0, 1) = 0$. However, equilibrium quantities such as $1/\sqrt{g}, g^{\rho \theta}$ become singular as we approach the magnetic axis. To avoid this behavior, we do not include the $\rho_{\rm{s}} = 0$ point exactly in our collocation grid. Instead, we impose the Dirichlet boundary condition at $\rho_{\rm{s}} = \epsilon$ where $\epsilon$ is a small number such that $\epsilon \in [10^{-3}, 10^{-2}]$. The value of $\epsilon$ does not affect our results for a small enough $\epsilon$.

For the kinetic energy matrix $\mathbb{B}$, these boundary conditions are applied before Cholesky decomposition by zeroing out the off-diagonal coupling terms in the block-diagonal blocks containing $\displaystyle \xi^{\rho_s}_0$ and $\displaystyle \xi^{\rho_s}_{N-1}$ shown in figure~(\ref{fig:Node-major}). However, we keep the diagonal $\displaystyle \xi^{\rho_s}_{0}$ and $\displaystyle \xi^{\rho_s}_{N-1}$ terms in diagonal blocks to ensure that they do not affect the calculation. After transforming the problem to~\eqref{eqn:standard-EVP}, we apply the boundary condition on the transformed potential energy matrix $\hat{\mathbb{A}}$ by eliminating the rows corresponding to nodes at $\rho_s = \epsilon$ and $\rho_s = 1$. Note that applying the Dirichlet boundary condition to $v^{\rho_s}$ is equivalent to applying the boundary condition to $\xi^{\rho_s}$. 

\subsection{Eigenvalue accuracy}
\label{sec:eigenvalue-accuracy}

The transformed matrix $\hat{\mathbb{A}}$ has a wide spectral range and contains eigenvalues close to zero, making the matrix poorly conditioned. Consequently, eigenvalues near the marginal stability boundary may lose relative accuracy in double precision. The wide spectral range results from both the spectral discretization and the structure of the ideal-MHD spectrum. The ideal-MHD force operator possesses continuous spectra, including the Alfv\'en and slow continua, and discrete eigenvalue sequences can accumulate at cluster points associated with these continua. In the spectra considered here, these accumulation points lie near marginal stability or on the stable side, $\lambda \leq 0$, whereas the unstable modes of interest, $\lambda > 0$, are discrete and separated from the accumulation point.~\texttt{AGNI} is designed to identify and optimize against these discrete unstable modes rather than resolve the stable spectrum or the ideal-MHD continua.

Although the stable cluster is not itself a target of the calculation, the full spectral range sets the finite-precision scale relevant to the unstable eigenvalues. For a backward-stable Hermitian eigensolver, the nominal absolute roundoff scale is $\mathcal{O}(\epsilon_{\mathrm{round}} \lvert \lvert \hat{\mathbb{A}} \lvert \lvert_2)$, where $\epsilon_{\mathrm{round}} \approx 10^{-16}$ is the unit-roundoff error. An eigenvalue close to the marginal-stability boundary should therefore not be regarded as numerically resolved when $\lvert \lambda \rvert \lesssim \epsilon_{\mathrm{round}} \lvert \lvert \hat{\mathbb{A}} \rvert \rvert_2$. On the other hand, an isolated unstable eigenvalue satisfying $\lvert \lambda \rvert \gg \epsilon_{\mathrm{round}} \lVert \hat{\mathbb{A}} \rVert_2$ can be clearly distinguished from marginal stability in double precision. For a Hermitian matrix, $\lvert \lvert \hat{\mathbb{A}} \rvert \rvert_2 = \mathrm{max}_{i} |\lambda_i|$.

\begin{figure}[h]
\centering
    \includegraphics[width=0.38 \textwidth, trim={2mm 2mm 1mm 1mm}, clip]{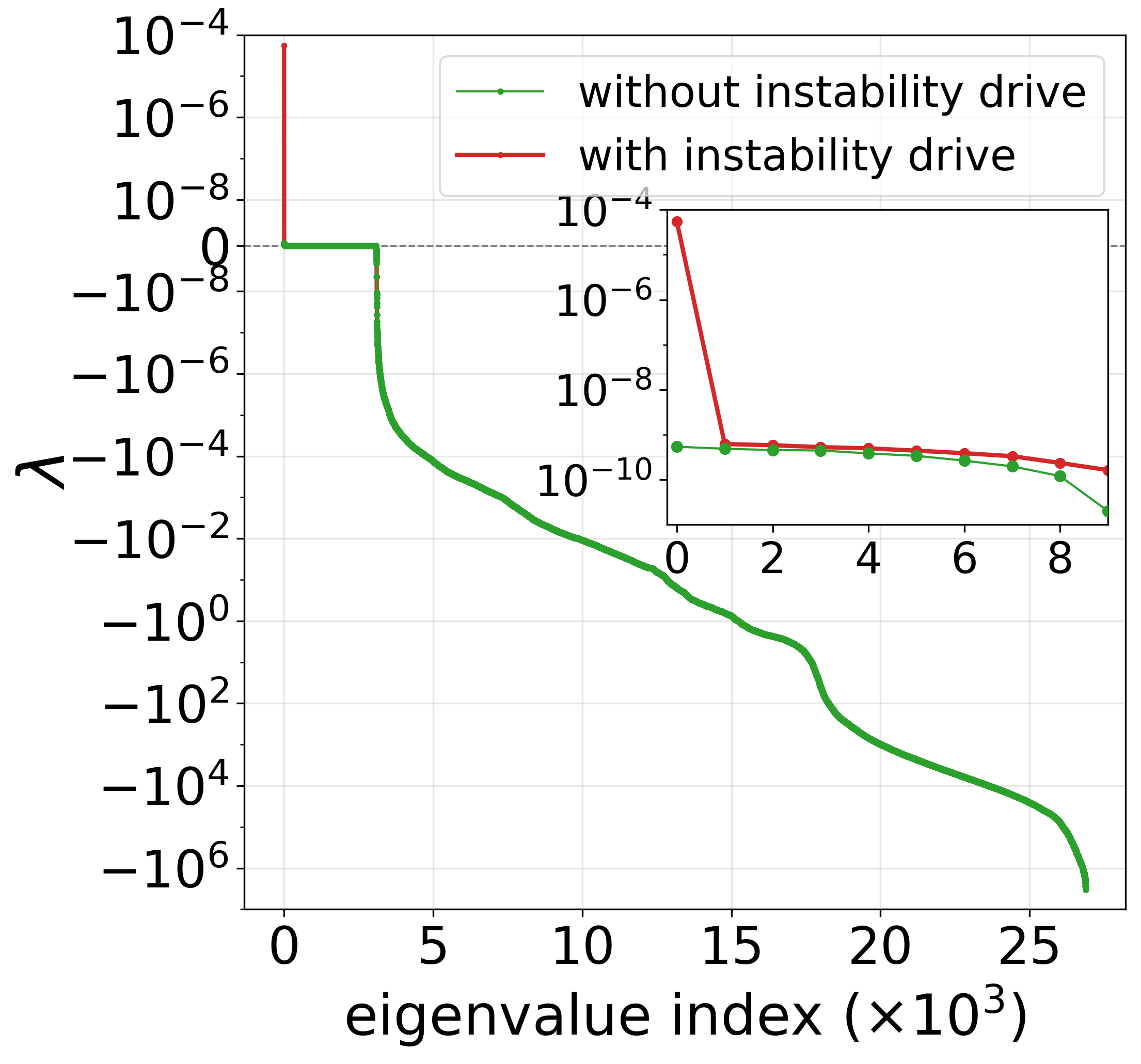}
    \caption{Full eigenspectrum of a stellarator equilibrium with and without the instability drive term $F$. Without the drive, the largest computed eigenvalues lie within the nominal roundoff scale and are numerically indistinguishable from zero. With the drive included, an isolated positive eigenvalue appears well above this scale.
    }
\label{fig:Instability-drive-test}
\end{figure}

For a typical tokamak or stellarator equilibrium, $\lVert \hat{\mathbb{A}} \rVert_2 \sim \mathcal{O}(10^6)$, giving a nominal roundoff error of $\mathcal{O}(10^{-10})$.
We demonstrate the wide spectral range and the difficulty to resolve close-to-marginal eigenvalues in figure~\ref{fig:Instability-drive-test} by analyzing the eigenspectrum of a modified Landreman, Buller, and Drevlak QH~\cite{Landreman2022QuasisymmetricBootstrap} stellarator equilibrium. The equilibrium is described in more detail in section~\ref{sec:benchmarks}. In figure $2$, we plot the eigenspectrum with and without the instability drive term $F$. 
Without the instability drive term, all eigenvalues $\lambda \leq 4 \cdot 10^{-10}$. The wide spectral range combined with double precision increases the eigenvalue uncertainty to $\mathcal{O}(10^{-10})$ which is where the most unstable eigenvalue is located for the case without any instability drive. Since the most unstable mode with instability drive is much larger than the eigenvalue uncertainty, it is the only truly resolved unstable mode.

\subsection{Shift-invert technique with Lanczos iterations}
To avoid computing the densely clustered stable and continuum-dominated portions of the spectrum, $\texttt{AGNI}$ uses the shift-invert method. Instead of solving the original problem, we solve the transformed problem
\begin{equation}
{\mathbb{H}_{\sigma}}^{-1}\bm{v} = \mu \bm{v}, \qquad \mathbb{H}_{\sigma} \equiv (\sigma \mathbb{I} - \hat{\mathbb{A}}),\,  \mu \equiv \frac{1}{\sigma - \lambda},
\end{equation}
so eigenvalues closest to the selected shift $\sigma$ become the largest-magnitude eigenvalues of the transformed operator. By choosing $\sigma$ near the unstable portion of the spectrum, the most unstable (and largest) eigenvalue $\mu$ can be extracted in a few iterations without calculating the complete eigenspectrum\footnote{Note that for a given mode, the shift-inverted operator has the same set of eigenfunctions as the unshifted operator, \textit{i.e.}, only the eigenvalue changes after the shift-invert operation.}. This substantially reduces both the computational cost and memory requirements relative to a dense eigendecomposition and becomes especially advantageous in the context of eigenvalue gradient calculation and gradient-based optimization. In particular, a full decomposition produces and stores a $3N \times 3N$ matrix containing all eigenvectors, whereas the iterative method stores only a small Krylov basis and the few requested eigenvectors.
The iterative Lanczos method for solving the shift-inverted eigenvalue problem is implemented in~\texttt{matfree}~\cite{kraemer2024gradients}, a~\texttt{jax} package. A shift-inverted operator is implemented in~\texttt{AGNI} separately using $\texttt{jax.scipy.linalg.lu\_solve}$. Both of these packages can be run on CPUs and GPUs. CPUs allow us to solve high-resolution problems (as they typically have more memory), whereas GPUs allow a lower resolution, but are at least an order of magnitude faster (massively multi-threaded) for both eigenvalue and gradient evaluation, suitable for gradient-based optimization. This effect is quantified in section~\ref{subsec:Convergence-study}.

Below, we present the pseudocode explaining the algorithm behind the default eigenvalue solver in~\texttt{AGNI}.

\par
\noindent\rule{\linewidth}{0.8pt}
\vspace{-2.8em}
\begin{center}
\captionof{algorithm}{Dense shift-invert Lanczos eigensolver}
\label{alg:shift_invert_lanczos}
\vspace{-0.8em}
\noindent\rule{\linewidth}{0.4pt}
\begin{algorithmic}[1]
\Require Equilibrium parameters $\mathbf{x}$; shift $\sigma$, not an eigenvalue of $\hat{\mathbb{A}}$, placed above $\lambda_{\max}$; Lanczos dimension $n_{\rm{mv}}$; start vector $\bm{q}_0$
\Ensure Most unstable eigenvector $\bm{v}$ and eigenvalue $\lambda$
\State Assemble the reduced matrix $\hat{\mathbb{A}}(\mathbf{x})$ and form $\mathbb{H}_{\sigma}=\sigma\mathbb{I}-\hat{\mathbb{A}}$
\State Factorize $\mathbb{H}_{\sigma}$ once: LU for any admissible $\sigma$, or Cholesky when $\sigma>\lambda_{\max}$
\State $\bm{q}_0 \gets \bm{q}_0/\lVert\bm{q}_0\rVert$
\State Apply $n_{\rm{mv}}$ steps of fully reorthogonalized Lanczos to $\bm{b}\mapsto\mathbb{H}_{\sigma}^{-1}\bm{b}$, reusing the stored factors, giving Ritz pairs $(\mu_\ell,\bm{v}_\ell)$
\State $\ell^{\star}\gets\arg\max_{\ell}\lvert\mu_{\ell}\rvert$
\State \Return $\bm{v}\gets\bm{v}_{\ell^{\star}}$, \quad $\lambda\gets\sigma-1/\mu_{\ell^{\star}}$
\end{algorithmic}
\end{center}
\vspace{-1.0em}
\noindent\rule{\linewidth}{0.4pt}

The shift $\sigma$ is user-defined, with a default value of $\sigma = 10^{-3}$. It can also be estimated from a low-resolution calculation of the eigenspectrum in
the preprocessing step. The Lanczos dimension is $n_{\rm{mv}}=50$ for the dense Lanczos LU (LLU) runs. 

As shown in section~\ref{subsec:Convergence-study}, this approach is computationally fast on a GPU as long as the dense reduced potential-energy matrix $\hat{\mathbb{A}}$ can be materialized and the shifted matrix $\mathbb{H}_{\sigma}$ can be factorized. The memory required for these dense matrix operations eventually limits the resolution that can be used on both CPUs and GPUs. For the hardware used here, this occurs around $N = 20000$ (problem size approximately $3N \times 3N$). We therefore replace the shift-invert Lanczos method described above with a matrix-free Jacobi--Davidson~\cite{sleijpen1996jacobi} method.

\subsection{Preconditioned matrix-free Jacobi--Davidson solver}
\label{subsec:matrix-free}
In this section, we will present how the matrix-free formulation performs operations with the matrix $\hat{\mathbb{A}}$, without assembling it. The eigensolver is the
Jacobi--Davidson method of algorithm~\ref{alg:jacobi_davidson} which replaces the shift-invert method, and the last part of this section constructs the preconditioner that its correction equation requires.
The obvious matrix-free route would be to retain the shift-invert Lanczos method and apply $\mathbb{H}_{\sigma}^{-1}$ iteratively. When the shift is larger than the largest eigenvalue of $\hat{\mathbb{A}}$, $\mathbb{H}_{\sigma}$ is Hermitian positive definite and the linear system
\begin{equation}
    \mathbb{H}_{\sigma}\bm{x}=\bm{b}
    \label{eqn:matrix-free-linear-system}
\end{equation}
can be solved, in principle, iteratively using the conjugate-gradient (CG) method. However, the number of iterations $m_{\rm{CG}}$ required to achieve convergence with a CG method scales approximately as $m_{\rm{CG}} = \mathcal{O}(\sqrt{\kappa(\mathbb{H}_{\sigma})})$~\cite{MEINARDUS1963}, where the condition number
\begin{equation}
    \kappa(\mathbb{H}_{\sigma}) = \frac{\sigma-\lambda_{\rm{min}}}{\sigma-\lambda_{\rm{max}}},
    \label{eqn:H-condition}
\end{equation}
where $\lambda_{\rm{min}} = \min(\lambda), \lambda_{\rm{max}} = \max(\lambda)$. For a shifted operator, at medium resolution of $24 \times 32 \times 12$, the condition number $\kappa(\mathbb{H}_{\sigma}) = \mathcal{O}(10^{10})$ which implies that to invert $\mathbb{H}_{\sigma}$, we would require a large number of conjugate gradient iterations $m_{\rm{CG}} = \mathcal{O}(10^5)$. The large condition number combined with the number of Lanczos iterations $n_{\rm{mv}} = \mathcal{O}(10^2)$ makes CG unsuitable for fast eigenvalue calculation and optimization. 

The Jacobi--Davidson method avoids the inversion entirely. Given the Ritz pair $(\lambda,\bm{u})$ and its residual $\bm{r}=\hat{\mathbb{A}}\bm{u}-\lambda\bm{u}$, the search space is enlarged by a vector $\bm{t}\perp\bm{u}$ ($\perp$ means orthogonal) satisfying the correction equation
\begin{equation}
    \mathbb{P}^{\dagger}\,\mathbb{H}_{\sigma}\,\mathbb{P}\,\bm{t} = \bm{r},
    \qquad \mathbb{P} = \mathbb{I}-\bm{u}\bm{u}^{\dagger}.
    \label{eqn:jd-correction}
\end{equation}
The projection removes the direction $\bm{u}$, along which $\mathbb{H}_{\sigma}$ is nearly singular by construction, so the smallest eigenvalue of the projected operator is $\sigma-\lambda_2$ rather than $\sigma-\lambda_{\max}$, where $\tilde{\lambda}_2$ is the second largest eigenvalue. The operator is positive definite on the complement of $\bm{u}$ whenever $\sigma>\lambda_{\max}$, so conjugate gradients apply. 
We fix a budget of $m_{\rm{in}}=200$ inner iterations rather than solving to a tolerance. The iteration also terminates early if $\operatorname{Re}(\bm{p}^{\dagger}\mathbb{H}_{\sigma}\bm{p})\leq0$ for a conjugate-gradient search direction $\bm{p}$, which cannot occur for $\sigma>\lambda_{\max}$ and therefore signals an inadmissible shift.

A preconditioner is still needed to make those $m_{\rm{in}}$ iterations effective. For a preconditioner $\mathbb{M}$ that preserves positive
definiteness, the preconditioned conjugate-gradient (PCG) problem becomes
\begin{equation}
     (\mathbb{P}^{\dagger}\mathbb{M}^{-1}\mathbb{P})(\mathbb{P}^{\dagger}\mathbb{H}_{\sigma} \mathbb{P})\bm{t}=(\mathbb{P}^{\dagger} \mathbb{M}^{-1} \mathbb{P}) \bm{r}.
    \label{eqn:matrix-free-linear-system2}
\end{equation}
An ideal preconditioner $\mathbb{M}$ would be a matrix that is computationally efficient to assemble and invert while closely approximating $\mathbb{H}_{\sigma}$. Such a preconditioner will cause the eigenvalues of the conditioned operator to cluster tightly, reducing the condition number, and accelerating convergence. To this end, we use a block-Jacobi preconditioner constructed from poloidal rings of the shifted operator. For each radial and toroidal index $(i,k)$, define the set
\begin{equation}
    \mathcal{R}_{ik}
    =
    \left\{(i,j,k):j=0,\ldots,N_{\theta}-1\right\},
    \label{eqn:ring-index}
\end{equation}
including all three components of the displacement at these collocation
points. The ring preconditioner is then
\begin{equation}
    \mathbb{M}_{\rm{ring}}
    =
    \operatorname{blockdiag}_{i,k}
    \left[
    \mathbb{H}_{\sigma}
    [\mathcal{R}_{ik},\mathcal{R}_{ik}]
    \right].
    \label{eqn:ring-preconditioner}
\end{equation}
Each block therefore contains the complete coupling between the three displacement components and all poloidal collocation points at fixed
$(\rho_i,\zeta_k)$, while coupling between different rings is omitted from the preconditioner. Constructing these blocks does not require materializing
$\hat{\mathbb{A}}$. As shown in section~\ref{sec:discretization}, the discretized terms in the energy integral have the general form
\begin{equation}
    \mathbb{T} = D_{x^{'}}^{\dagger}\mathrm{diag}(W  \sqrt{g}\, \mathrm{E})D_{x}.
\end{equation}
For any ring index set $\mathcal{R}$, the corresponding principal submatrix can therefore be assembled directly as
\begin{equation}
    \mathbb{T}[\mathcal{R},\mathcal{R}] = D_{x'}[:,\mathcal{R}]^{\dagger} \mathrm{diag}(W  \sqrt{g}\, \mathrm{E}) D_{x}[:,\mathcal{R}].
    \label{eqn:restricted-assembly}
\end{equation}
Consequently, the same discretized energy integral used to apply $\hat{\mathbb{A}}$ matrix-free is used to construct the ring blocks, without
forming the full matrix. The ring blocks are Cholesky decomposed independently, and applying $\mathbb{M}_{\rm{ring}}^{-1}$ consists of independent triangular solves of matrices of size $3 \nt \times 3 \nt$ for each ring because of the block-diagonal structure. This technique is similar to the one used to simplify the kinetic energy matrix in section~\ref{subsec:efficient-Cholesky}. The ring preconditioner reduces the condition number by around two orders of magnitude, from $\mathcal{O}(10^{10})$ to $\mathcal{O}(10^8)$.

However, the smallest eigenvalues of the preconditioned shifted operator still limit PCG
convergence. 
We further accelerate the linear solve with an additive coarse-space correction~\cite{tang2009comparison}, built from a matrix $\mathbb{Z}$ of $k_{\rm{defl}}$ vectors transferred from a lower-resolution discretization to the fine grid,
\begin{equation}
    \mathbb{M}^{-1} = \mathbb{M}_{\rm{ring}}^{-1} + \mathbb{Z}\left(\mathbb{Z}^{\dagger}\mathbb{H}_{\sigma}\mathbb{Z}\right)^{+}\mathbb{Z}^{\dagger},
    \label{eqn:deflated-preconditioner}
\end{equation}
where $^+$ is a pseudoinverse. The coarse vectors are chosen to approximate the eigenvectors associated with the smallest eigenvalues of a coarse ring-preconditioned operator by solving a preconditioned dense generalized eigenvalue problem on a coarse grid
\begin{equation}
    {\mathbb{H}}^{\rm{coarse}}_{\sigma}\bm{z}_i = \eta_i {\mathbb{M}}^{\rm{coarse}}_{\rm{ring}}\bm{z}_i,
    \label{eqn:coarse-deflation}
\end{equation}
retain the eigenvectors corresponding to the smallest $\eta_i$, interpolate them to the fine grid, and use them to construct $\mathbb{Z}$. The interpolated eigenvector corresponding to the smallest $\eta_i$ also provides the start vector $\bm{u}_0$ of algorithm~\ref{alg:jacobi_davidson} on the first call. In subsequent calls during an optimization, the eigenvector from the previous evaluation is used instead. Each application of $\mathbb{M}^{-1}$ in the correction equation then consists of independent triangular solves for the ring blocks together with one coarse solve. The coarse space $\mathbb{Z}$ is constructed once per eigensolve, from the unprojected operator $\mathbb{H}_{\sigma}$, whereas the correction equation~\eqref{eqn:jd-correction} is solved with $\mathbb{P}^{\dagger}\mathbb{H}_{\sigma}\mathbb{P}$ and preconditioned by $\mathbb{P}^{\dagger}\mathbb{M}^{-1}\mathbb{P}$, with $\mathbb{P}$ changing at every outer iteration. The algorithm is explained in more detail in algorithm~\ref{alg:jacobi_davidson}.

\par\medskip
\noindent\rule{\linewidth}{0.8pt}
\vspace{-2.8em}
\begin{center}
\captionof{algorithm}{Matrix-free Jacobi--Davidson eigensolver}
\label{alg:jacobi_davidson}
\vspace{-0.8em}
\noindent\rule{\linewidth}{0.4pt}
\begin{algorithmic}[1]
\Require Matrix-free product $\bm{x}\mapsto\hat{\mathbb{A}}\bm{x}$; preconditioner $\mathbb{M}^{-1}$ from~\eqref{eqn:deflated-preconditioner}; start vector $\bm{u}_0$ from~\eqref{eqn:coarse-deflation}; parameters $\sigma$, $m_{\max}$, $k_{\rm{keep}}$, $m_{\rm{in}}$, $n_{\rm{out}}$, $\tau_{\lambda}$ of table~\ref{tab:jd-parameters}
\Ensure Most unstable eigenvector $\bm{v}$ and eigenvalue $\lambda$
\Statex \textit{The columns of $\mathbb{V}$ are the search directions $\bm{v}_1,\ldots,\bm{v}_j$ and the columns of $\mathbb{W}$ are $\hat{\mathbb{A}}\bm{v}_1,\ldots,\hat{\mathbb{A}}\bm{v}_j$.}
\State $\bm{v}_1\gets\bm{u}_0/\lVert\bm{u}_0\rVert$, \ $\bm{w}_1\gets\hat{\mathbb{A}}\bm{v}_1$, \ $j\gets1$, \ $\lambda_{\rm{prev}}\gets\infty$
\For{$i=1,\ldots,n_{\rm{out}}$}
  \Statex \quad\textit{Rayleigh--Ritz on $\operatorname{span}(\bm{v}_1,\ldots,\bm{v}_j)$}
  \State $\mathbb{S}\gets\tfrac{1}{2}\bigl(\mathbb{V}^{\dagger}\mathbb{W}+(\mathbb{V}^{\dagger}\mathbb{W})^{\dagger}\bigr)$ \Comment{$j\times j$, Hermitian}
  \State Diagonalize $\mathbb{S}$, giving $(\lambda_{\ell},\bm{y}_{\ell})$ with $\lambda_1$ largest
  \State $\bm{u}\gets\mathbb{V}\bm{y}_1$, \ $\hat{\mathbb{A}}\bm{u}\gets\mathbb{W}\bm{y}_1$, both normalized by $\lVert\bm{u}\rVert$ \Comment{no new product}
  \State $\bm{r}\gets\hat{\mathbb{A}}\bm{u}-\lambda_1\bm{u}$, \ $\delta\gets\lvert \lambda_1-\lambda_{\rm{prev}}\rvert/\lvert \lambda_1\rvert$, \ $\lambda_{\rm{prev}}\gets\lambda_1$
  \Statex \quad\textit{Correction equation, solved approximately}
  \State Apply at most $m_{\rm{in}}$ PCG iterations to~\eqref{eqn:jd-correction}, preconditioned by $\mathbb{P}\mathbb{M}^{-1}\mathbb{P}$, giving $\bm{t}\perp\bm{u}$; stop early if $\operatorname{Re}(\bm{p}^{\dagger}\mathbb{H}_{\sigma}\bm{p})\leq0$
  \Statex \quad\textit{Restart if the search space is full}
  \If{$j=m_{\max}$}
    \State $\bm{v}_{\ell}\gets\mathbb{V}\bm{y}_{\ell}$, $\bm{w}_{\ell}\gets\mathbb{W}\bm{y}_{\ell}$ for $\ell=1,\ldots,k_{\rm{keep}}$; \ $j\gets k_{\rm{keep}}$
  \EndIf
  \Statex \quad\textit{Expand the search space}
  \State $\bm{t}\gets\bm{t}-\mathbb{V}(\mathbb{V}^{\dagger}\bm{t})$, applied twice
  \If{$\lVert\bm{t}\rVert$ is finite and nonzero}
    \State $\bm{v}_{j+1}\gets\bm{t}/\lVert\bm{t}\rVert$, \ $\bm{w}_{j+1}\gets\hat{\mathbb{A}}\bm{v}_{j+1}$, \ $j\gets j+1$ \Comment{one product}
  \EndIf
  \If{$\delta\leq\tau_{\lambda}$} \textbf{break} \EndIf
\EndFor
\State Repeat lines 3--6 on the final $\bm{v}_1,\ldots,\bm{v}_j$
\State \Return $\bm{v}\gets\bm{u}$, \ $\lambda\gets \lambda_1$
\end{algorithmic}
\end{center}
\vspace{-3.0mm}
\noindent\rule{\linewidth}{0.4pt}

\begin{table}[h]
\centering
\caption{Parameters of the Jacobi--Davidson eigensolver and the default
values used in this work.}
\label{tab:jd-parameters}
\begin{tabular}{llc}
\hline
Symbol & Meaning & Default \\
\hline
$\sigma$          & shift, must exceed $\lambda_{\max}$          & $10^{-3}$ \\
$m_{\max}$        & search directions stored before a restart    & 60 \\
$k_{\rm{keep}}$   & Ritz vectors retained at a restart           & 10 \\
$m_{\rm{in}}$     & PCG iterations per correction solve          & 200 \\
$n_{\rm{out}}$    & maximum outer iterations                     & 500 \\
$\tau_{\lambda}$   & relative tolerance on $\lambda_1$             & $10^{-8}$ \\
$k_{\rm{defl}}$   & coarse-correction vectors in $\mathbb{Z}$    & 50 \\
$N_{\rm{c}}$      & resolution of the coarse problem             & 11520 \\
\hline
\end{tabular}
\end{table}

In the next section, we will explain how the time and memory costs scale with matrix size for different eigenvalue solvers and how these scalings justify the use of matrix-free solvers in our work.

\subsection{Cost of the eigensolvers}
\label{subsec:solver-choice}
Table~\ref{tab:table-2} compares the cost of the two~\texttt{AGNI} eigensolvers with that of the dense eigensolvers in~\texttt{scipy} and~\texttt{jax}. The four dense methods share the $\mathcal{O}(N^2)$ memory and $\mathcal{O}(N^3)$ time of the factorization or diagonalization, but~\texttt{eigsh} and LLU compute only the $k$ requested eigenpairs and are therefore much faster in practice than~\texttt{eigh} (section~\ref{subsec:Convergence-study}). The JD method stores $\mathcal{O}(m_{\Sigma}N)$ on the fine grid, and its run time is set by the number of PCG iterations, i.e., by the quality of the preconditioner. Its coarse level, however, is solved densely: for $N_{\rm{c}} = 11\,520$ the coarse matrix alone occupies about $9.6$~GB, which dominates the memory of the JD runs in table~\ref{tab:table-3}.
 
\begin{table*}[h]
\centering
\caption{Comparison of eigensolvers. $N$ is the total number of collocation points, so $\mathrm{dim}(\hat{\mathbb{A}}) \approx 3N$, $k$ is the number of requested eigenpairs, and $N_{\rm{c}}$ is the number of coarse-grid points.}
\label{tab:eigensolvers}
\small
\setlength{\tabcolsep}{4pt}
\begin{tabular}{lccccc}
\hline
Solver & Device & Spectrum & $\mathrm{max}(\mathrm{dim}(\hat{\mathbb{A}}))$ & Memory & Time \\
\hline
\texttt{eigh (scipy)} & CPU & full & $\leq 1\times10^{5}$ & $\mathcal{O}(N^2)$ & $\mathcal{O}(N^3)$ \\
\texttt{eigsh (scipy)} & CPU & top $k$ & $\leq 1\times10^{5}$ & $\mathcal{O}(N^2+n_{\rm{mv}}N)$ & $\mathcal{O}(N^3+n_{\rm{mv}}N^2)$ \\
\texttt{eigh (jax.linalg)} & CPU/GPU & full & $\leq 3\times10^{4}$ & $\mathcal{O}(N^2)$ & $\mathcal{O}(N^3)$ \\
Lanczos\! \&\! dense LU & CPU/GPU & top $k$ & $\leq 6\times10^{4}$ & $\mathcal{O}(N^2+n_{\rm{mv}}N)$ & $\mathcal{O}(N^3+n_{\rm{mv}}N^2)$ \\
matrix-free JD & CPU/GPU & top $1$ & $9\times10^{4}$\,$^{a}$ & $\mathcal{O}(m_{\Sigma}N + N_{\rm{c}}^2)$ & $\mathcal{O}(n_{\rm{out}}m_{\rm{in}}C_{\rm{mv}} + N_{\rm{c}}^3)$ \\
\hline
\end{tabular}
 
\vspace{0.5em}
\begin{minipage}{0.95\textwidth}
\footnotesize
Each JD outer iteration performs one Rayleigh--Ritz extraction on at most $m_{\max}$ vectors and at most $m_{\rm{in}}$ PCG iterations, each costing one matrix--vector product $C_{\rm{mv}}$. $m_{\Sigma} = 2m_{\max} + k_{\rm{defl}} + 9\nt$ counts the stored values per collocation point: the search space and its image under $\hat{\mathbb{A}}$, the coarse vectors, and the Cholesky factors of the ring blocks. The $N_{\rm{c}}$ terms are the dense coarse solve of~\eqref{eqn:coarse-deflation}. $^{a}$Largest size tested (table~\ref{tab:table-3}).
\end{minipage}
\label{tab:table-2}
\end{table*}

In the following section, we will use these methods to calculate and compare eigenmodes and eigenvalues, over different resolution values and hardware.
\section{Benchmarks and tests}
\label{sec:benchmarks}

We benchmark~\texttt{AGNI} against~\texttt{NIMSTELL} \cite{sovinec_semi-implicit_2026}, a nonlinear initial-value code that solves the non-ideal MHD model given by equations \eqref{eqn:MHD-continuity}-\eqref{eqn:div-B}. \texttt{NIMSTELL} uses a modified form of~\eqref{eqn:MHD-induction} to incorporate divergence cleaning~\cite{sovinec2004}. The time-dependent fields in \texttt{NIMSTELL} are decomposed into steady-state and perturbed parts, and nonlinear terms are retained in general. However, the linear stability of a given MHD equilibrium, which is treated as the steady state, can be modeled by excluding the nonlinear terms. In such calculations, the perturbation becomes exponentially larger along the fastest growing eigenmode than along other eigenmodes, which allows the eigenfunction and growth rate to be inferred.

\begin{figure}[h]
    \centering
    \begin{subfigure}[b]{0.5\textwidth}
    \centering
        \includegraphics[width=0.48 \textwidth, trim={2mm 2mm 0mm 0mm}, clip]{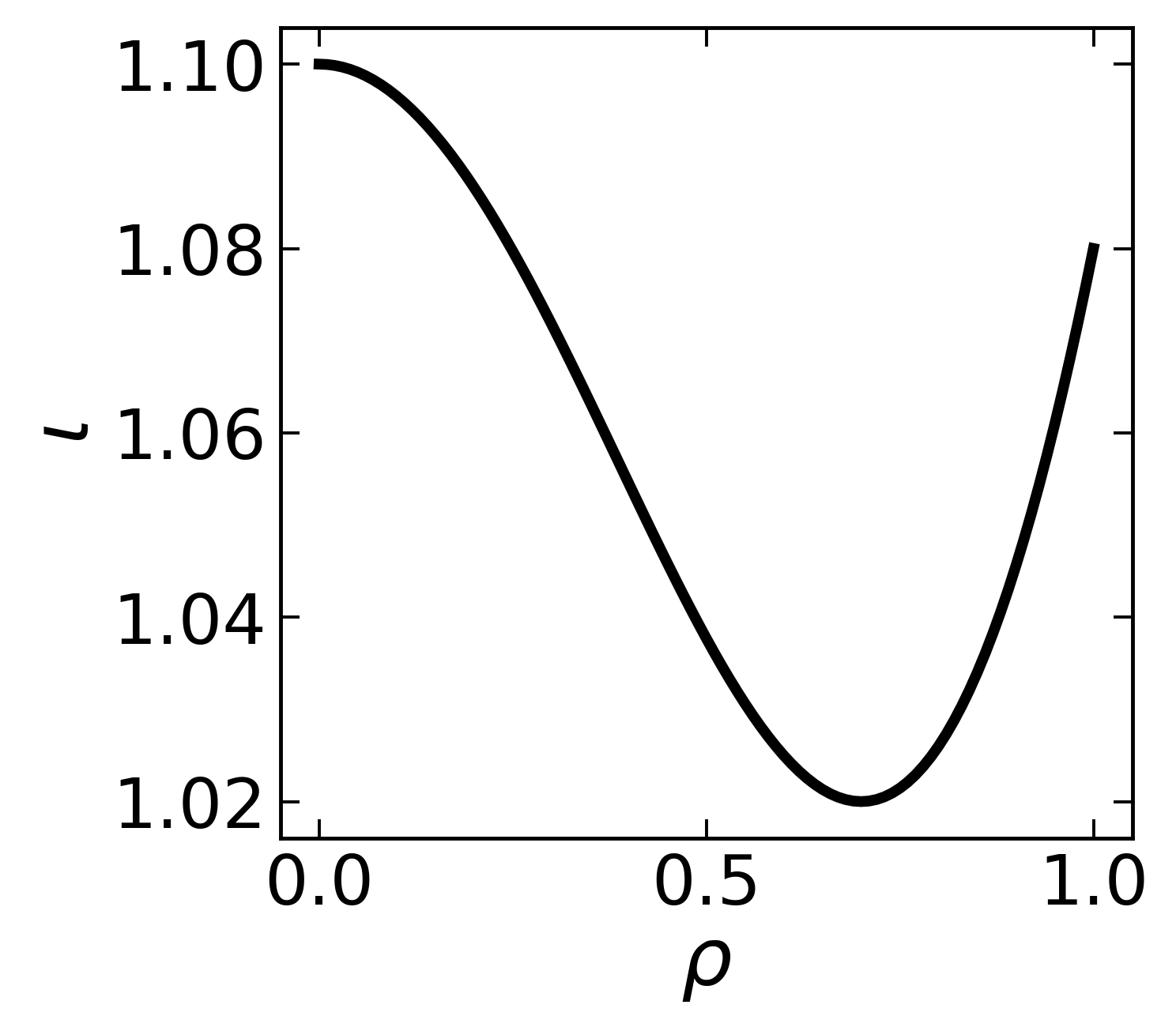}
        \hspace*{-3mm}
        \includegraphics[width=0.5 \textwidth, trim={2mm 2mm 0mm 0mm}, clip]{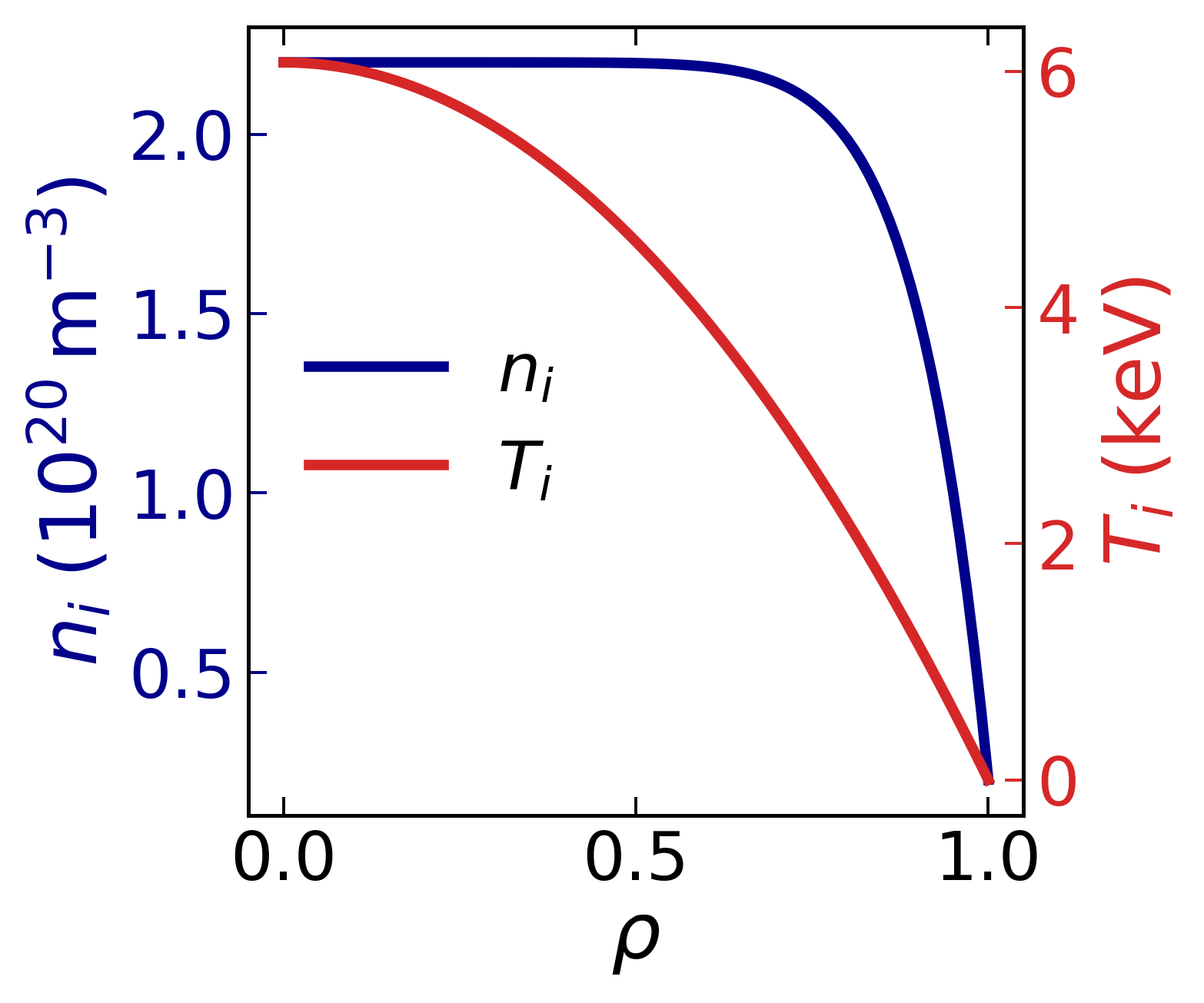}
        \caption{Modified QH stellarator}
    \label{fig:LBD-QH}
    \end{subfigure}
    \hspace*{-3mm}
    \begin{subfigure}[b]{0.5\textwidth}
    \centering
        \includegraphics[width=0.49 \textwidth, trim={0mm 2mm 0mm 0mm}, clip]{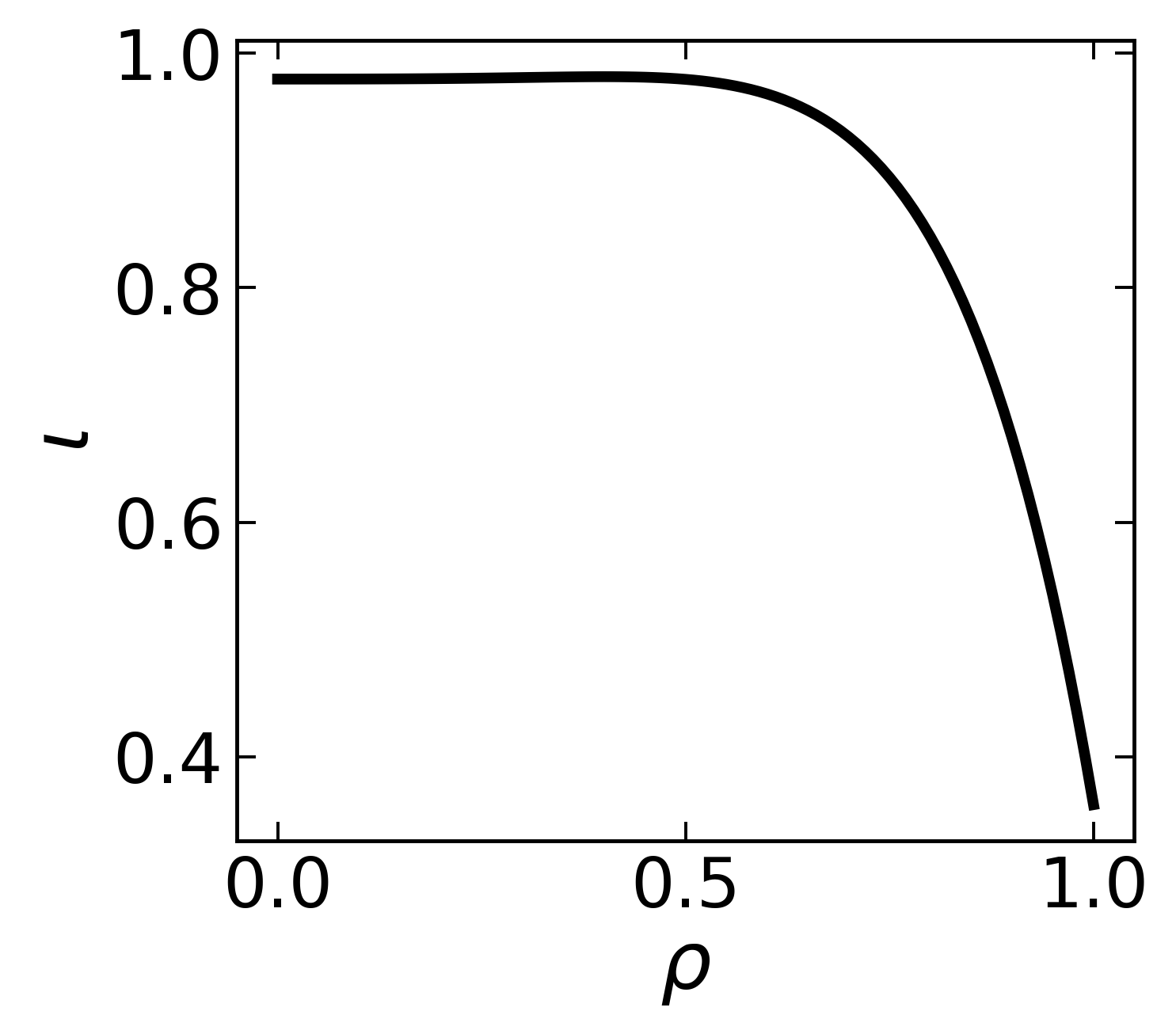}
        \hspace*{-3mm}
        \includegraphics[width=0.49 \textwidth, trim={0mm 2mm 0mm 0mm}, clip]{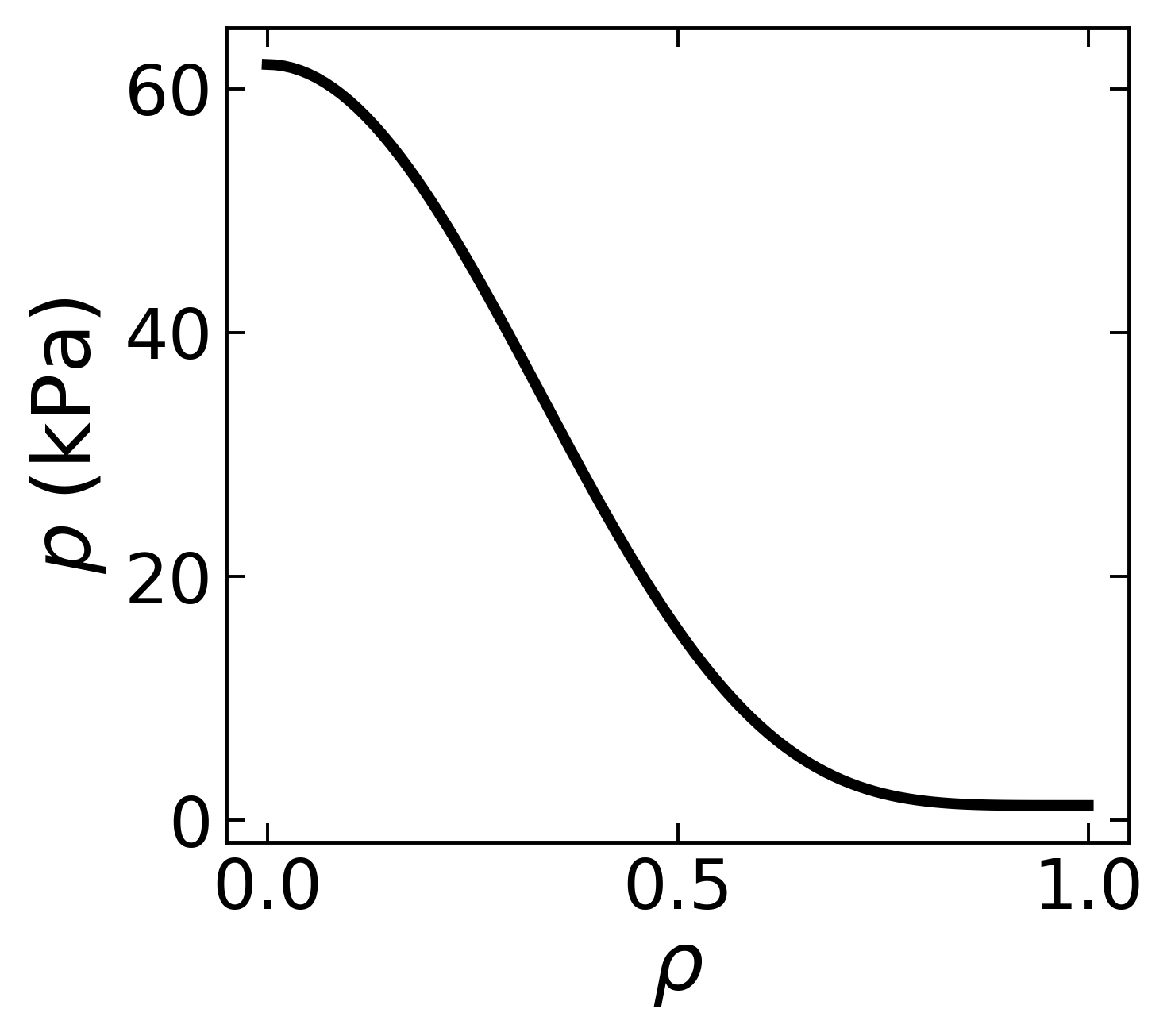}
        \caption{DSHAPE tokamak}
    \label{fig:DSHAPE}
    \end{subfigure}
\caption{Rotational transform, ion density $n_i$ and ion temperature $T_i$ profiles of the modified LBD QH stellarator (left) and Rotational Transform and pressure profiles for the DSHAPE tokamak (right). The electron density $n_e = n_i$ and temperature $T_e = T_i$.}
\label{fig:Profiles}
\end{figure}

For our benchmark, we choose two equilibria: a quasi-helically (QH) symmetric equilibrium from Landreman--Buller--Drevlak (LBD)~\cite{Landreman2022QuasisymmetricBootstrap} with a modified rotational transform profile and a DSHAPE tokamak. The profiles are shown in figure~\ref{fig:Profiles}. 
Both equilibria are unstable to near-resonant, ideal pressure-driven modes similar to those outlined in Ref. \cite{patil_verication_nodate},  where the~\texttt{NIMSTELL} results are verified against CASTOR3D for the QH equilibrium at $\beta=2.5\%$ and $\beta=5.0\%$.

In the present calculations, the viscous stress tensor $\bm{\Pi}$ is assumed to be isotropic,
\begin{equation}
    \bm{\Pi} = -M n_0 \nu_{\rm{iso}} \left( \bm{\nabla V} + (\bm{\nabla V})^T - \frac{2}{3} \bm{I} (\bm{\nabla \cdot V}) \right) ,
\end{equation}
where $n_0$ is a constant density parameter and $\nu_{\rm{iso}}$ is the viscous diffusivity. The electrical resistivity $\eta$ is assumed to be constant and uniform, whereas the thermal conductivity tensor $\bm{\kappa}$ is set to zero. Additionally, \texttt{NIMSTELL}'s numerical-interchange stabilization method is applied~\cite{sovinec_semi-implicit_2026}. 
At the plasma boundary, the no-flux, no-slip and Dirichlet boundary conditions are used for the density, velocity, and temperature, respectively. The perturbed magnetic field at the boundary is constrained to be tangential, \emph{i.e.}, a perfectly conducting wall is assumed.


\subsection{Stellarator benchmark}
For the LBD QH benchmark case, both~\texttt{NIMSTELL} and~\texttt{AGNI} solve for the eigenmode in the $n=\{0, 4, 8, \ldots\}$ family of toroidal Fourier harmonics\footnote{Since the $n=0$ family is field-period symmetric, we only solve the stability problem in a single field period which reduces the number of grid points $N_{\zeta}$, making the computation more efficient. For non-field period symmetric modes $n \pmod \NFP \neq 0$, we must solve the problem in the full domain.}. 
\texttt{NIMSTELL} uses a spectral Fourier basis in the toroidal direction and a spectral-element mesh for discretizing physical quantities at each toroidal cross-section. The 2D spectral-element mesh includes two regions. The annular region consists of a polar mesh with 28 and 64 biquartic elements in the radial and poloidal directions, respectively. The central region near the magnetic axis hosts a logically rectangular $16\times16$-element mesh.
Since we model only one of the four field periods of the equilibrium, the toroidal variation of the geometry, equilibrium and perturbed fields is represented by Fourier series with mode numbers $n= \NFP k,~k=0,\ldots,10$ where $\NFP=4$ is the number of field periods. Fourier transforms use a grid of 64 poloidal planes within one field period. 
To check numerical convergence, we use 48 and 128 biquartic elements in the radial and poloidal directions of the annular mesh, respectively, and a $32\times32$ central mesh. The toroidal resolution is increased to $n=\NFP k,~k=0,\ldots,15$. The increase in spatial resolution changes the growth rate by only $0.2\%$, confirming convergence.

In Patil~\textit{et al.}~\cite{patil_verication_nodate}, the growth rate is independent of the viscous diffusivity and the electrical resistivity over the ranges $\nu_{\rm{iso}} \in [1.47\times10^{-5}, 1.47\times10^{-4}] ~\mathrm{m^2/s}$ and $\eta \in [1.85\times10^{-10}, 1.85\times10^{-6}]~\Omega\,\mathrm{m}$ for the QH cases. We therefore use $\nu_{\rm{iso}}=1.47\times10^{-5}~\mathrm{m^2/s}$ and $\eta = 1.85 \times 10^{-8}~\Omega\,\mathrm{m}$, for which the NIMSTELL model should closely resemble the ideal-MHD model of~\texttt{AGNI}.
We also set the parameter $n_0=2.2\times10^{20}~\text{m}^{-3}$, which is equal to the electron density on the magnetic axis.

\begin{figure}[h]
    \centering
    \begin{subfigure}[b]{0.502\textwidth}
    \centering
        \includegraphics[width=\textwidth, trim={14mm 0mm 0mm 0mm}, clip]{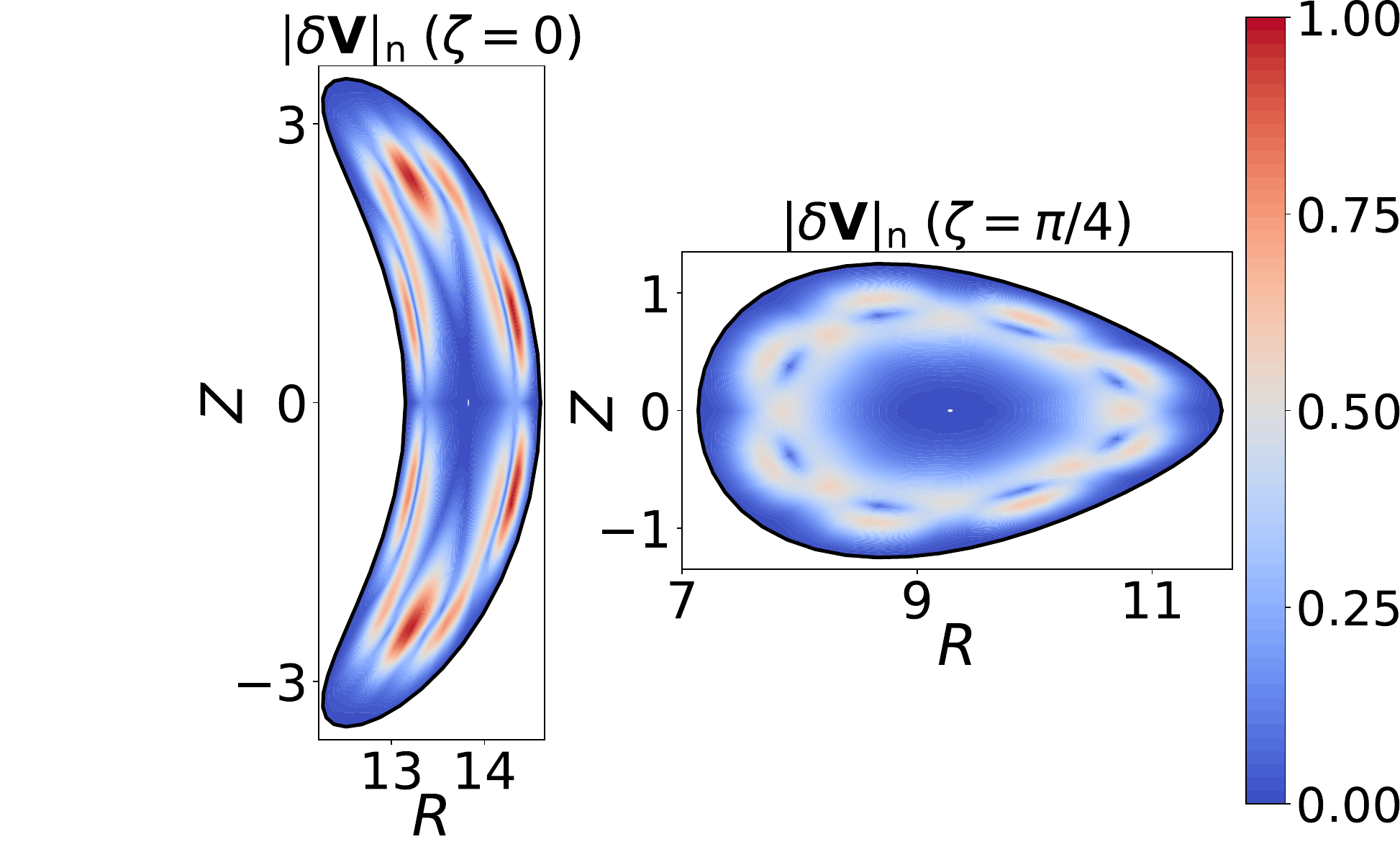}\\[2.5mm]
        \caption{~\texttt{AGNI}, $\gamma = 3.7\times 10^4 \mathrm{rad/s}$}
        \label{fig:AGNI}
    \end{subfigure}
    \,
    \begin{subfigure}[b]{0.475\textwidth}
        \centering
        \includegraphics[width=\textwidth, trim={10mm 8mm 10mm 14mm}, clip]{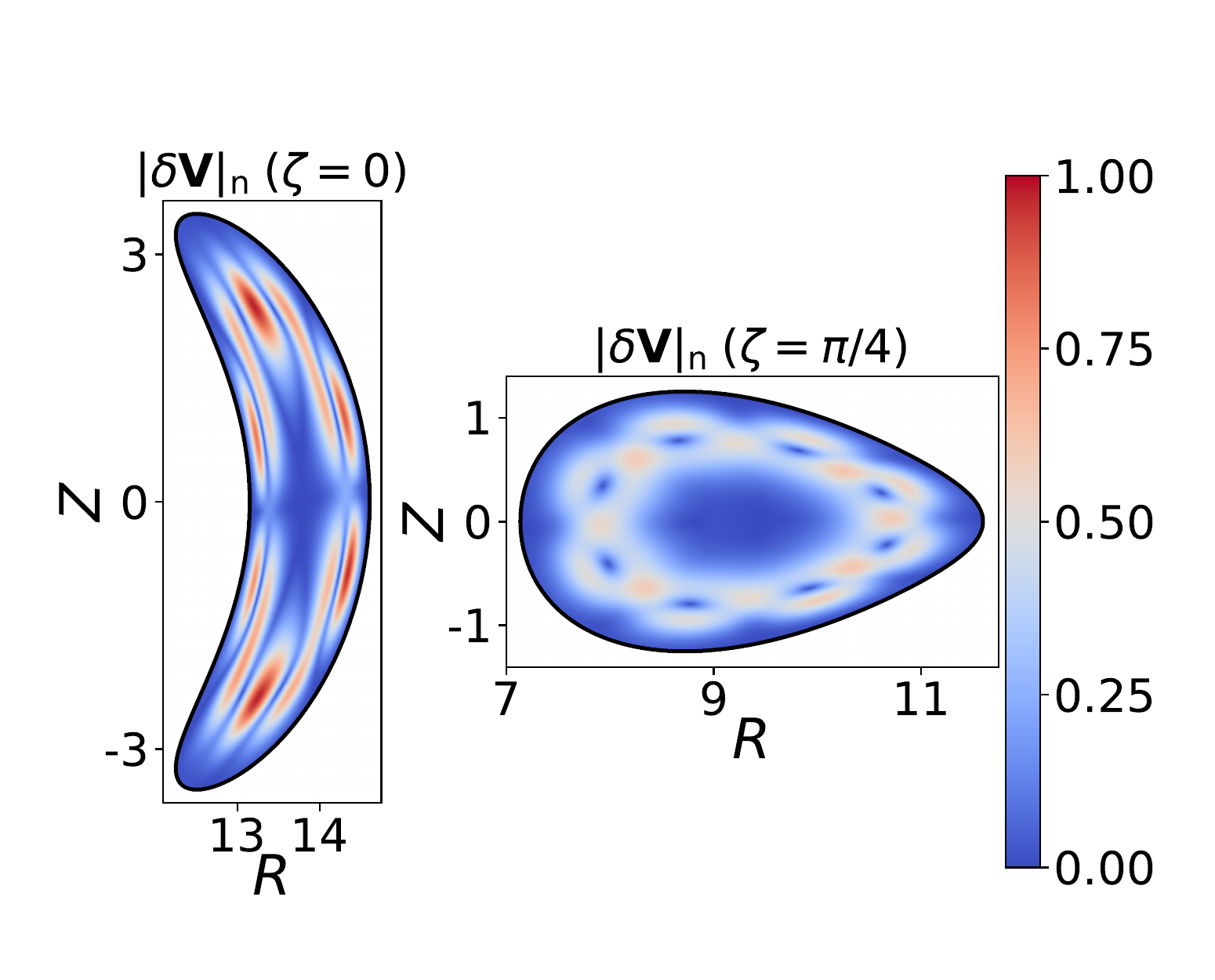}
        \caption{~\texttt{NIMSTELL}, $\gamma = 3.6\times 10^4 \mathrm{rad/s}$}
        \label{fig:NIMSTELL}
    \end{subfigure}
\caption{The eigenfunction is compared for the most unstable mode in ~\texttt{NIMSTELL} with ~\texttt{AGNI} for the modified LBD QH equilibrium. Note that for this study $\Gamma = 5/3$ in both the codes.}
\label{fig:AGNI-NIMSTELL-benchmark}
\end{figure}

We run~\texttt{AGNI} on a $48 \times 48 \times 16$ grid and limit the maximum poloidal mode number to $m = 8$ and toroidal mode number to $n=8$ when calculating the differentiation matrices. The mode truncation is necessary to filter out higher-$n$ modes and obtain a dominant mode that can be compared across the two codes. 
For comparison, we calculate the most unstable growth rate $\gamma = (\sqrt{\lambda}/a_{\rm{N}}) (B_{\rm{N}}/\sqrt{\mu_0 n_0 M})$ and perturbed speed of the plasma $\lvert \delta\! \bm{V}\rvert = ||\gamma| \tilde{\bm{\xi}}|$ (from $\delta \bm{V} = \partial \tilde{\bm{\xi}}/\partial t$) corresponding to that mode
\begin{equation}
    \lvert \delta\!\bm{V} \rvert = |\gamma| \sqrt{g_{\rho \rho}{\tilde{\xi^{\rho}}}^2 + g_{\theta \theta}{\tilde{\xi^{\theta}}}^2 + g_{\zeta \zeta}{\tilde{\xi^{\zeta}}}^2 + 2 (g_{\rho \theta} \tilde{\xi^{\rho}} \tilde{\xi^{\theta}} + g_{\theta \zeta} \tilde{\xi^{\theta}} \tilde{\xi^{\zeta}} +  g_{\zeta \rho} \tilde{\xi^{\zeta}} \tilde{\xi^{\rho}})}
    \label{eqn:perturbed-speed}
\end{equation}
and compare the normalized speed $\lvert \delta\!\bm{V} \rvert_{\mathrm{n}} = \lvert \delta\!\bm{V} \rvert/\mathrm{max}(\lvert \delta\!\bm{V}\rvert)$ at two different cross-sections in figure~\ref{fig:AGNI-NIMSTELL-benchmark}.
For the LBD equilibrium, $\lambda = 8.34 \times 10^{-5}, M = 1.67 \times 10^{-27} \rm{kg}, n_0 = 2.2 \times 10^{20} \rm{m}^{-3}, a_{\rm{N}} = 1.707\, \rm{m}, B_{\rm{N}} = 4.677\, \rm{T}$, gives $\gamma = 3.67 \times 10^{4}\, \rm{rad/s}$ which is within $2\%$ of the growth rate from~\texttt{NIMSTELL}. We find reasonable agreement between both the eigenfunctions and the eigenvalues. 

The dominant mode $m = n = 4$ can be clearly seen by the $8$ poloidal lobes in the $\zeta = \pi/4$ cross-section. The peak also occurs close to $\iota = 1.02$ where the magnetic shear $d\iota/d\rho = 0$ which indicates that this is most likely an interchange mode. 
The~\texttt{NIMSTELL} eigenfunction shows slight up-down asymmetry. In stellarator-symmetric equilibria, the eigenfunctions have definite parities under the point reflection about the origin, $(\theta, \zeta) \to (-\theta, -\zeta)$ \cite{strumberger2016,strumberger2019}.
If eigenfunctions with opposite parities have growth rates that are either degenerate, or very close to each other \cite{strumberger2019}, a superposition of the two eigenfunctions can arise in initial-value calculations. We expect this to be the cause of the up-down asymmetry in~\texttt{NIMSTELL}, since a specific parity is not imposed.

\subsection{Tokamak benchmark}
For the DSHAPE tokamak, the annular region of the~\texttt{NIMSTELL} mesh consists of a polar mesh with 28 and 32 biquartic elements in the radial and poloidal directions, respectively. The central region hosts a logically rectangular $8\times8$-element mesh. The toroidal Fourier transforms are based on $32$ poloidal planes.
Informed by the QH cases, we use $n_0=2\times10^{20} ~\mathrm{m}^{-3}$, $\nu_{\rm{iso}}=1\times10^{-5}~\mathrm{m^2/s}$, and $\eta = 1.26 \times 10^{-9}~\Omega\,\mathrm{m}$.

\begin{figure}[h]
    \centering
    \begin{subfigure}[b]{0.1817\textwidth}
    \centering
        \includegraphics[width=\textwidth, trim={8mm 0mm 0mm 0mm}, clip]{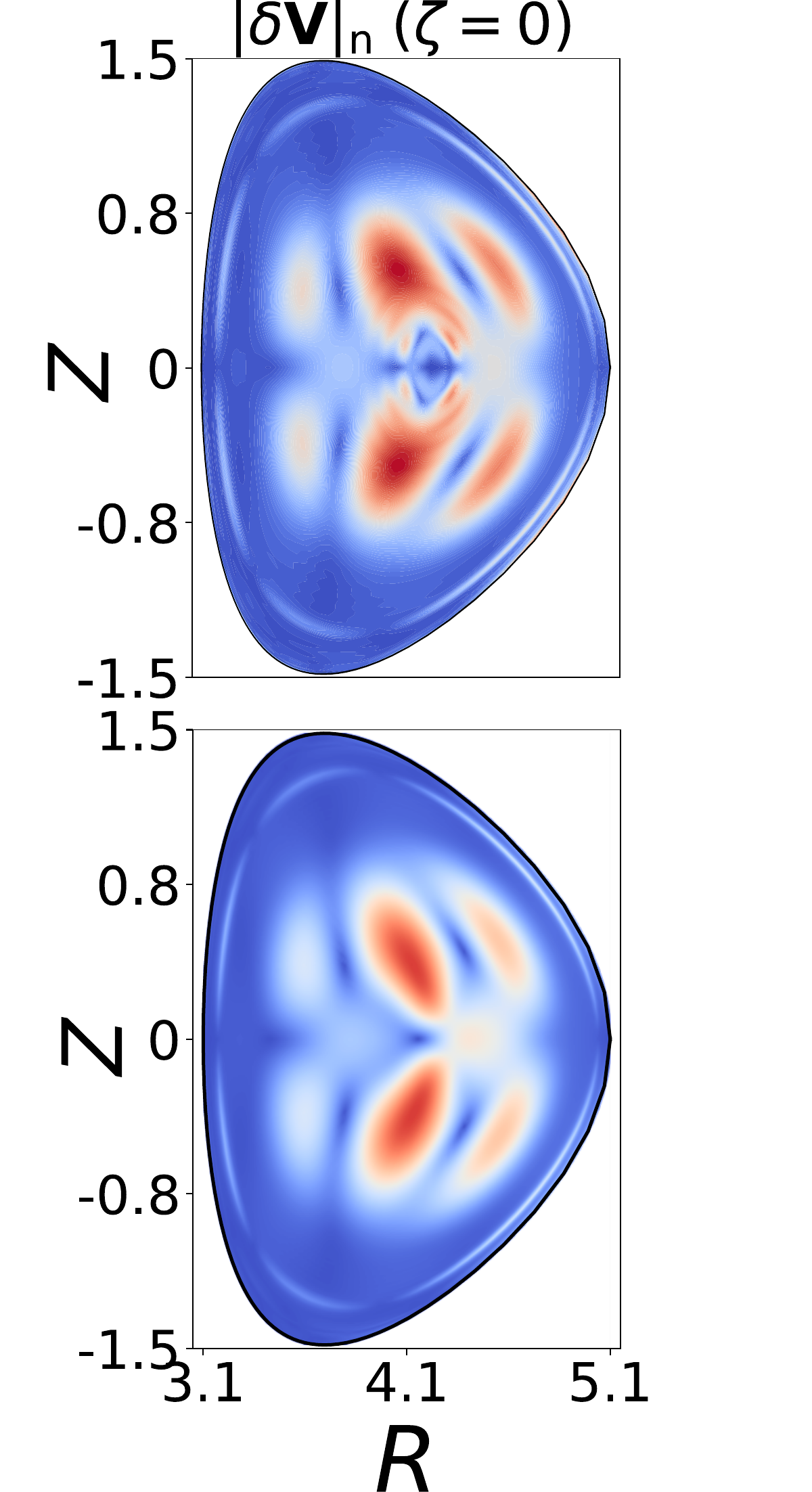}
        \hspace*{-3mm}\caption{$n=2$}
        \label{fig:n2}
    \end{subfigure}
    \hspace*{-5mm}
    \begin{subfigure}[b]{0.18\textwidth}
        \centering
        \includegraphics[width=\textwidth, trim={8mm 0mm 0mm 0mm}, clip]{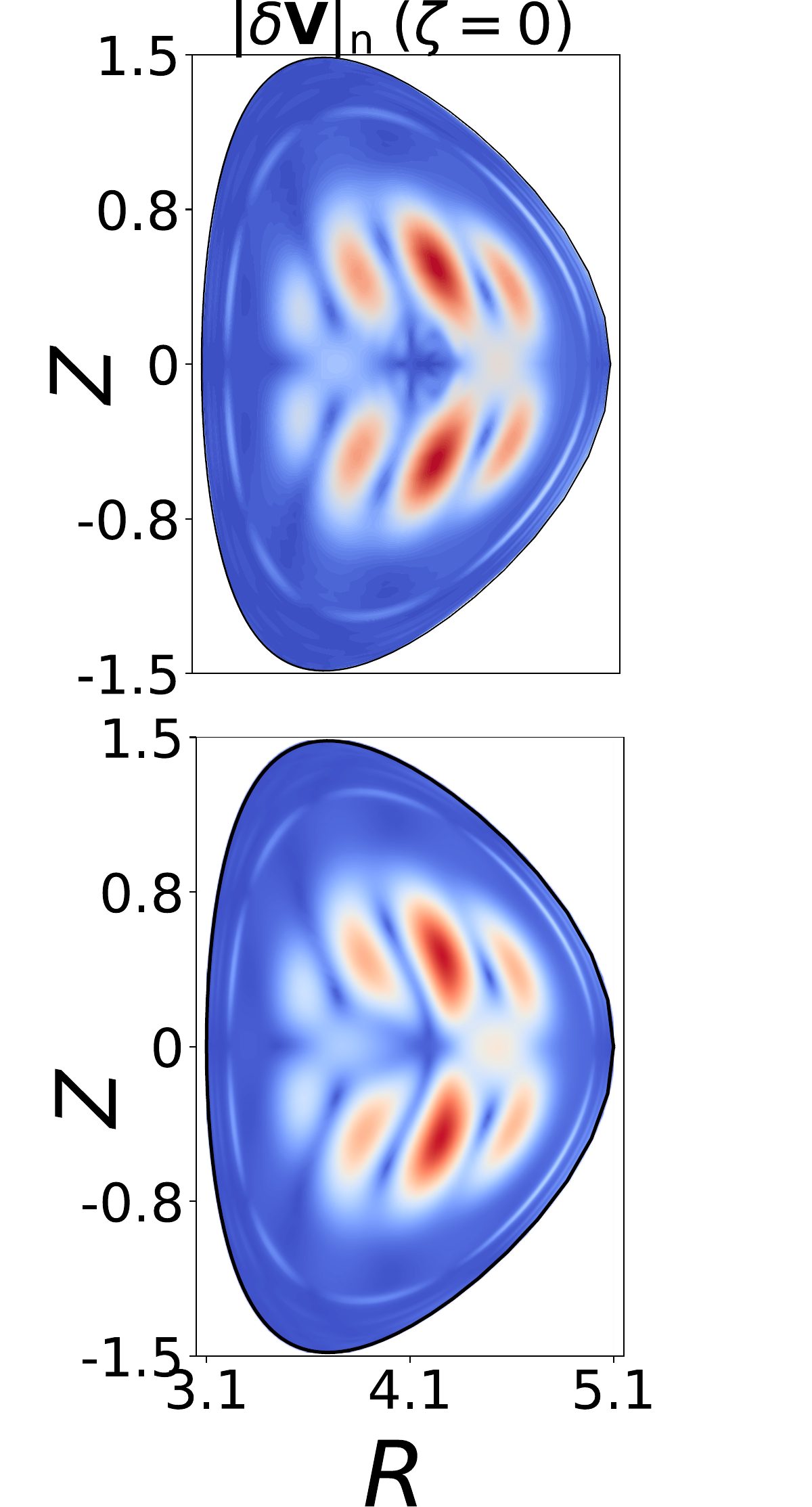}
        \caption{$n=3$}
        \label{fig:n3}
    \end{subfigure}
    \hspace*{-6mm}
    \begin{subfigure}[b]{0.1805\textwidth}
        \centering
        \includegraphics[width=\textwidth, trim={8mm 0mm 0mm 0mm}, clip]{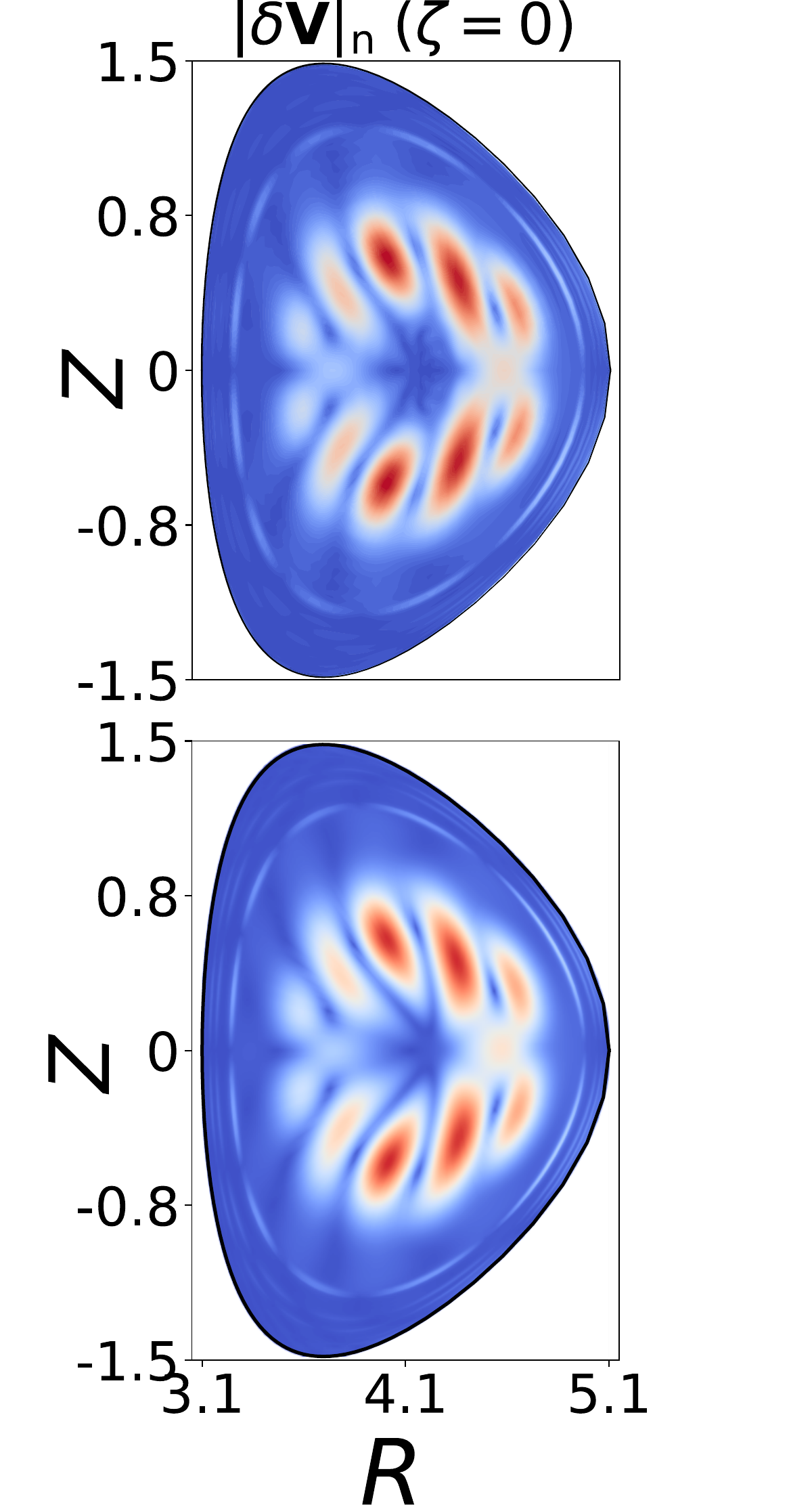}
        \caption{$n=4$}
        \label{fig:n4}
    \end{subfigure}
    \hspace*{-8mm}
    \begin{subfigure}[b]{0.187\textwidth}
        \centering
        \includegraphics[width=\textwidth, trim={19mm 0mm 6mm 0mm}, clip]{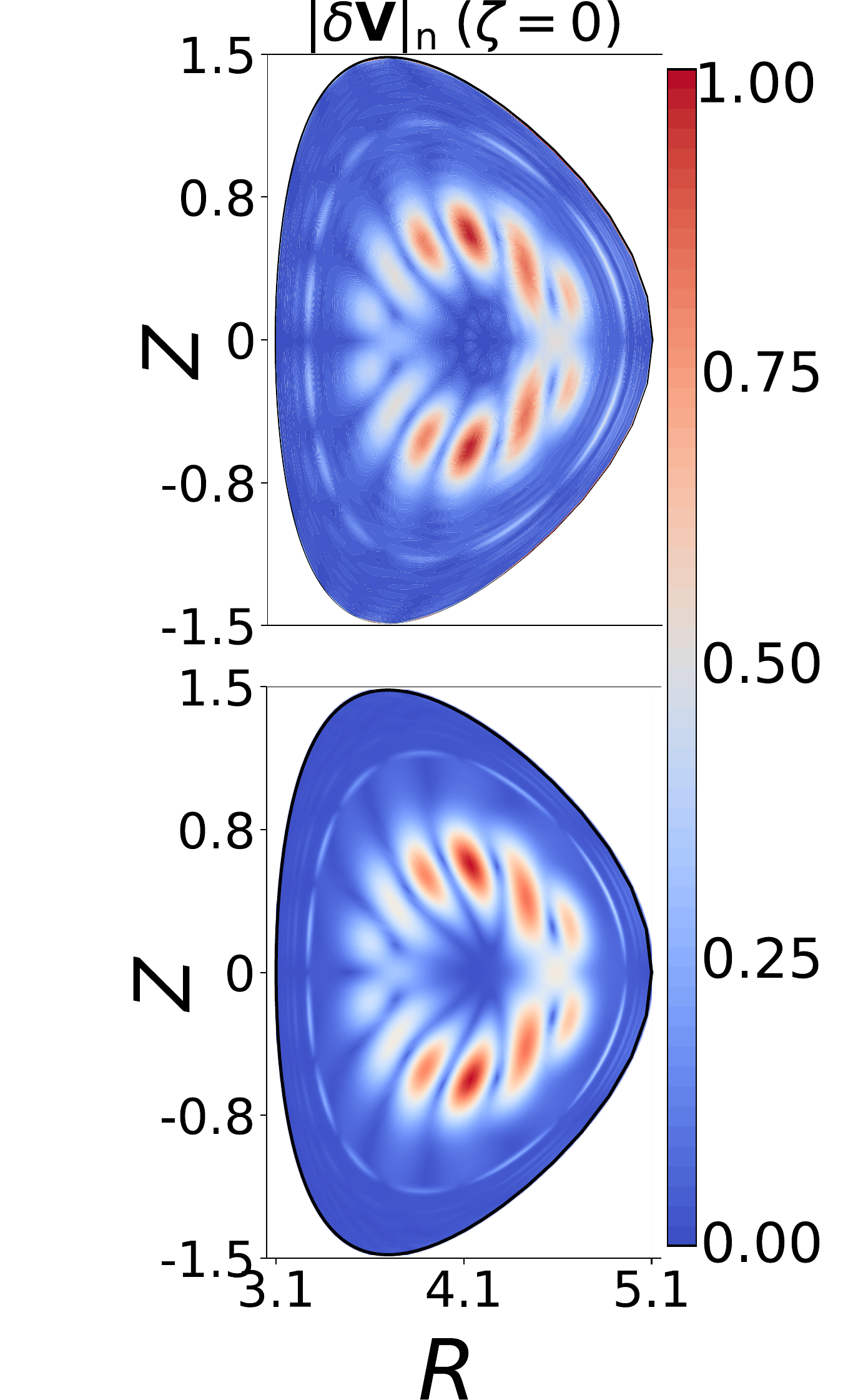}
        \caption{$n=5$}
        \label{fig:n6}
    \end{subfigure}
    \hspace*{-0mm}
    \begin{subfigure}[b]{0.22\textwidth}
        \centering
        \includegraphics[width=1.0\textwidth, trim={0mm 0mm 0mm 0mm}, clip]{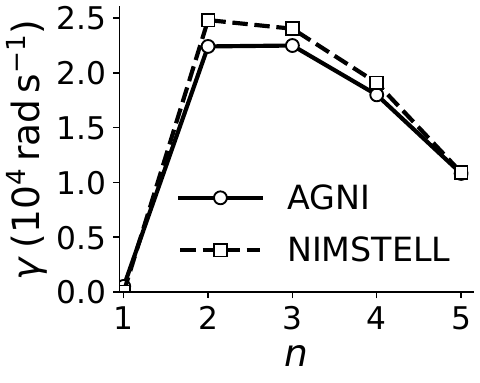}\\[10mm]
        \caption{$\gamma (n)$}
        \label{fig:gamma-vs-n}
    \end{subfigure}
\caption{Comparison of eigenfunctions between~\texttt{AGNI} (top row) and~\texttt{NIMSTELL} (bottom row) for the DSHAPE equilibrium along with the growth rate $\gamma$ trend with the toroidal mode number $n$. The $n=1$ mode was found to be stable in~\texttt{NIMSTELL} and weakly unstable in~\texttt{AGNI}. Note that for this study $\Gamma = 5/3$ in both the codes.}
\label{fig:AGNI-NIMSTELL-DSHAPE-benchmark}
\end{figure}

We discretize the energy integral in~\texttt{AGNI} using a Zernike basis on a $96 \times 96$ grid with the maximum poloidal mode limited to $4 n$ and solve for $n = 1$--$5$ modes separately and solve the eigenproblem on a GPU using the dense Lanczos method. We compare the perturbed speed~\eqref{eqn:perturbed-speed} and the growth rate trend with $n$ across the two codes in figure~\ref{fig:AGNI-NIMSTELL-DSHAPE-benchmark}.
We find that eigenfunctions, growth-rate values and growth rate trends match well with~\texttt{NIMSTELL}. The dominant mode peaks in the region of low shear $\rho \in [0, 0.5]$, with the dominant poloidal mode number $m = n$. Moreover, we can also reproduce the radially local feature close to the plasma boundary. Each mode took approximately $90$ seconds on an A100 GPU.


\subsection{Convergence study}
\label{subsec:Convergence-study}
To better understand the convergence and cost requirements of~\texttt{AGNI}, we run the stellarator benchmark case on both CPU and GPU at various grid resolutions while fixing the maximum poloidal and toroidal modes used to calculate the differentiation matrices. All the cases were run on the Perlmutter supercomputer with the CPU runs using a single node with two AMD EPYC 7713 (Milan) processors with a total of $128$ cores and 450 GB of memory, and GPU runs using a single NVIDIA A100 GPU with 80 GB of memory (VRAM). We present the eigenvalue calculation times in table~\ref{tab:table-3}.
%
%
%
%
%
\begin{table}[h]
    \centering
    \caption{Comparison of GPU and CPU eigenvalues and corresponding wall-clock times.}
    \label{tab:gpu_cpu_comparison}
    \begin{tabular}{c c c c c c c}
        \hline
        $N_{\rho} \times N_{\theta} \times N_{\zeta}$
        & GPU solver
        & Time (GPU)
        & Time (CPU)
        & $\lambda_{\mathrm{GPU}}$
        & $\lambda_{\mathrm{CPU}}$ \\
        \hline
        $8 \times 24 \times 8$
        & dense Lanczos
        & $11.3\,\mathrm{s}$
        & $12.7\,\mathrm{s}$
        & $1.45\times10^{-3}$
        & $1.45\times10^{-3}$ \\
        \hline

        $16 \times 32 \times 8$
        & dense Lanczos
        & $12.2\,\mathrm{s}$
        & $45.0\,\mathrm{s}$
        & $6.91\times10^{-5}$
        & $6.91\times10^{-5}$\\
        \hline

        $24 \times 40 \times 12$
        & dense Lanczos
        & $26.4\,\mathrm{s}$
        & $154.9\,\mathrm{s}$
        & $8.35\times10^{-5}$
        & $8.35\times10^{-5}$\\
        \hline


        $32 \times 48 \times 16$
        & JD
        & $120.8\,\mathrm{s}$
        & $1129.6\,\mathrm{s}$
        & $8.33\times10^{-5}$
        & $8.34\times10^{-5}$ \\
        \hline

        $40 \times 48 \times 16$
        & JD
        & $141.1\, \mathrm{s}$
        & $1667.5\,\mathrm{s}$
        & $8.34\times10^{-5}$
        & $8.34\times10^{-5}$ \\
        \hline
    \end{tabular}
\begin{minipage}{0.95\textwidth}
\footnotesize
$\sigma = 10^{-3}$ for all runs. JD parameters in table~\ref{tab:jd-parameters}. The CPU runs used~\texttt{scipy.sparse.linalg.eigsh}.
\end{minipage}
\label{tab:table-3}
\end{table}

We find that the GPU solver is faster than the CPU solver for the given choice of matrix-free parameters but the accuracy of the GPU run can suffer at high-resolution values so picking the appropriate set of parameters like $\sigma, m_{\rm{max}}, m_{\rm{in}}, n_{\rm{out}}$, and coarse deflation eigensolve resolution becomes crucial. Next, we calculate the gradient of the maximum eigenvalue with respect to all the equilibrium parameters on both CPU and GPU and time them. The timings are presented in table~\ref{tab:table-gradients}.
\begin{table}[h]
    \centering
    \caption{GPU and CPU eigenvalue gradients in reverse mode.}
    \label{tab:gpu_cpu_gradient_comparison}
    \begin{tabular}{c c c}
        \hline
        $N_{\rho} \times N_{\theta} \times N_{\zeta}$
        & Time (GPU)
        & Time (CPU)\\
        \hline
        
        $8 \times 24 \times 8$
        & $0.04\,\mathrm{s}$
        & $1.82\,\mathrm{s}$\\
        \hline

        $16 \times 32 \times 8$
        & $0.07\,\mathrm{s}$
        & $5.91\,\mathrm{s}$\\
        \hline
        
        $24 \times 40 \times 12$
        & $0.12\,\mathrm{s}$
        & $10.00\,\mathrm{s}$\\
        \hline

        
        $32 \times 48 \times 16$
        & $0.23 \,\mathrm{s}$
        & $34.90\,\mathrm{s}$\\
        \hline
        
        $40 \times 48 \times 16$
        & $0.26 \,\mathrm{s}$
        & $35.32 \,\mathrm{s}$\\
        \hline
    \end{tabular}
\begin{minipage}{0.95\textwidth}
\footnotesize
The gradient $\partial \lambda/\partial \bm{x}$ is calculated with respect to all the $7444$ equilibrium parameters to isolate the time taken by the gradient calculation from the set of physics constraints that are applied during optimization. We have also excluded the compilation time for all of these runs.
\end{minipage}
\label{tab:table-gradients}
\end{table}
A single~\texttt{AGNI} run and gradient calculation are much faster on accelerated hardware such as a GPU than on a CPU. As each optimization requires multiple calls to the eigenvalue solver and gradient calculator, GPUs are more suitable for a fast, gradient-based optimization framework.

In the next section, we will explore in detail the gradient calculation process using automatic differentiation and present related tests.

\subsection{Reverse-mode gradient calculation}
\begin{figure}[h]
    \centering
    \begin{subfigure}[b]{0.313\textwidth}
    \centering
        \includegraphics[width=\textwidth, trim={2mm 2mm 1mm 2mm}, clip]{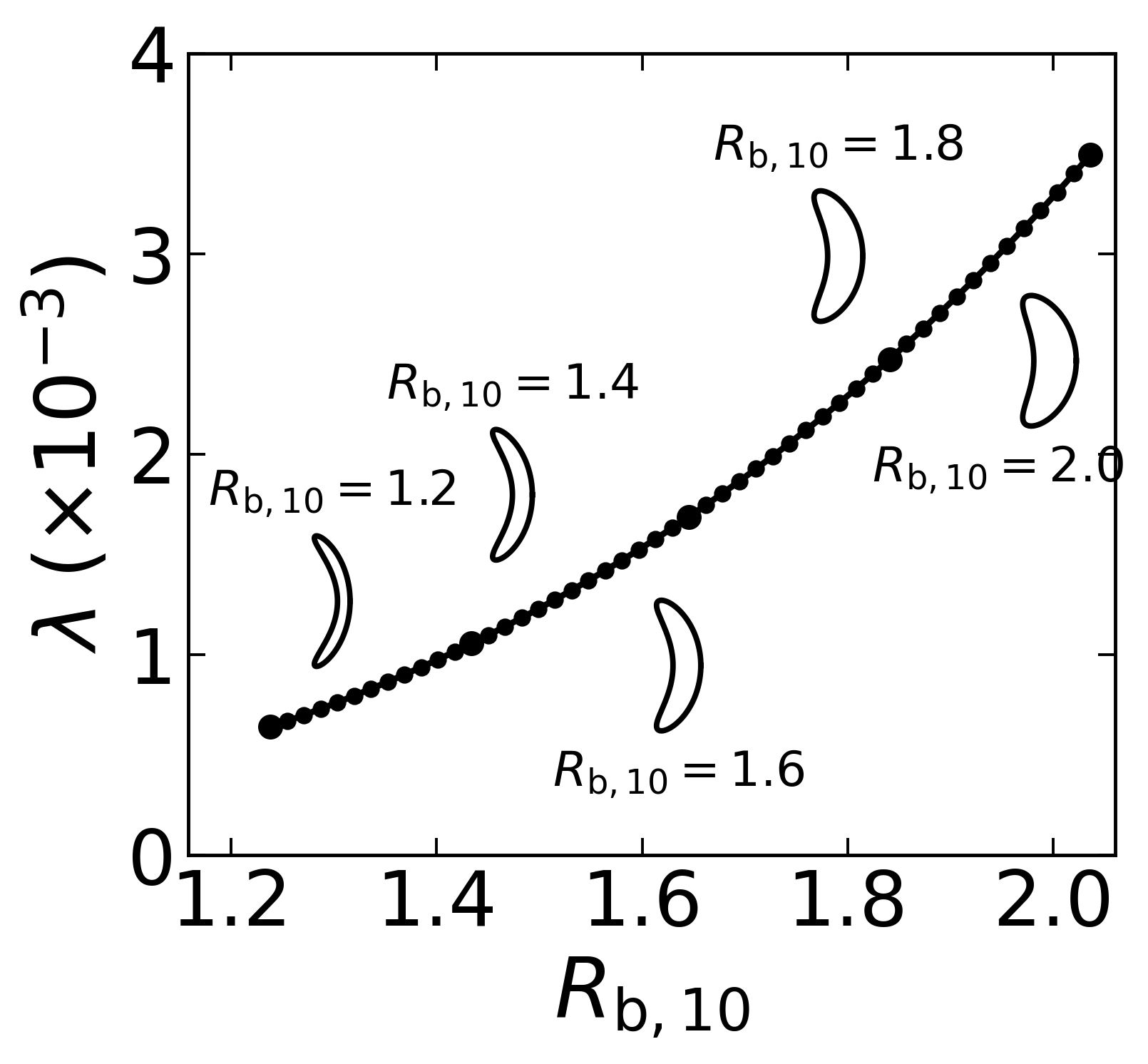}
        \caption{Eigenvalue landscape}
        \label{fig:landscape}
    \end{subfigure}
    \quad
    \begin{subfigure}[b]{0.30\textwidth}
        \centering
        \includegraphics[width=\textwidth, trim={2mm 2mm 2mm 2mm}, clip]{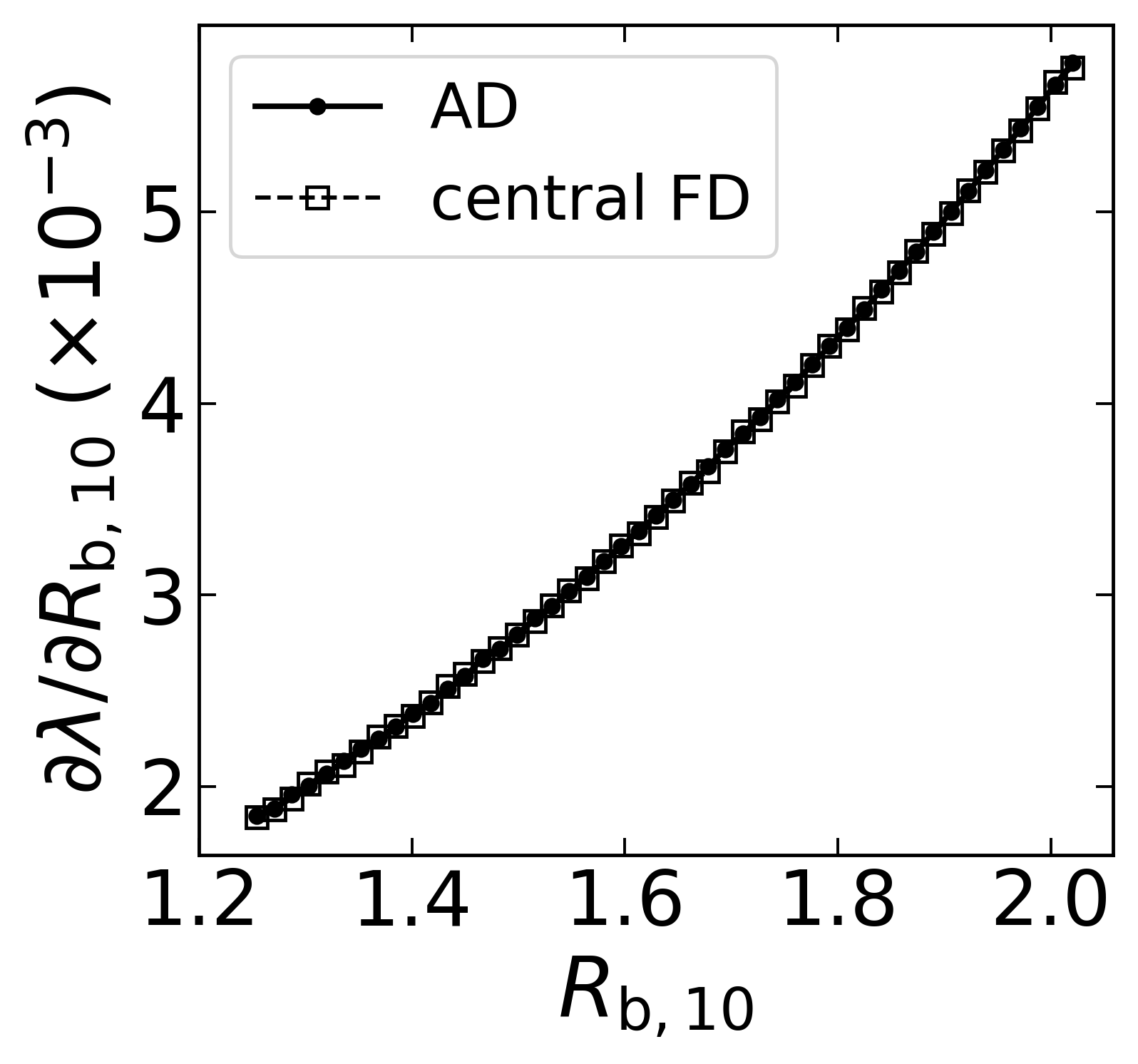}
        \caption{Gradient comparison}
        \label{fig:FD-vs-AD}
    \end{subfigure}
    \,
    \begin{subfigure}[b]{0.3115\textwidth}
        \centering
        \includegraphics[width=\textwidth, trim={2mm 2mm 2mm 2mm}, clip]{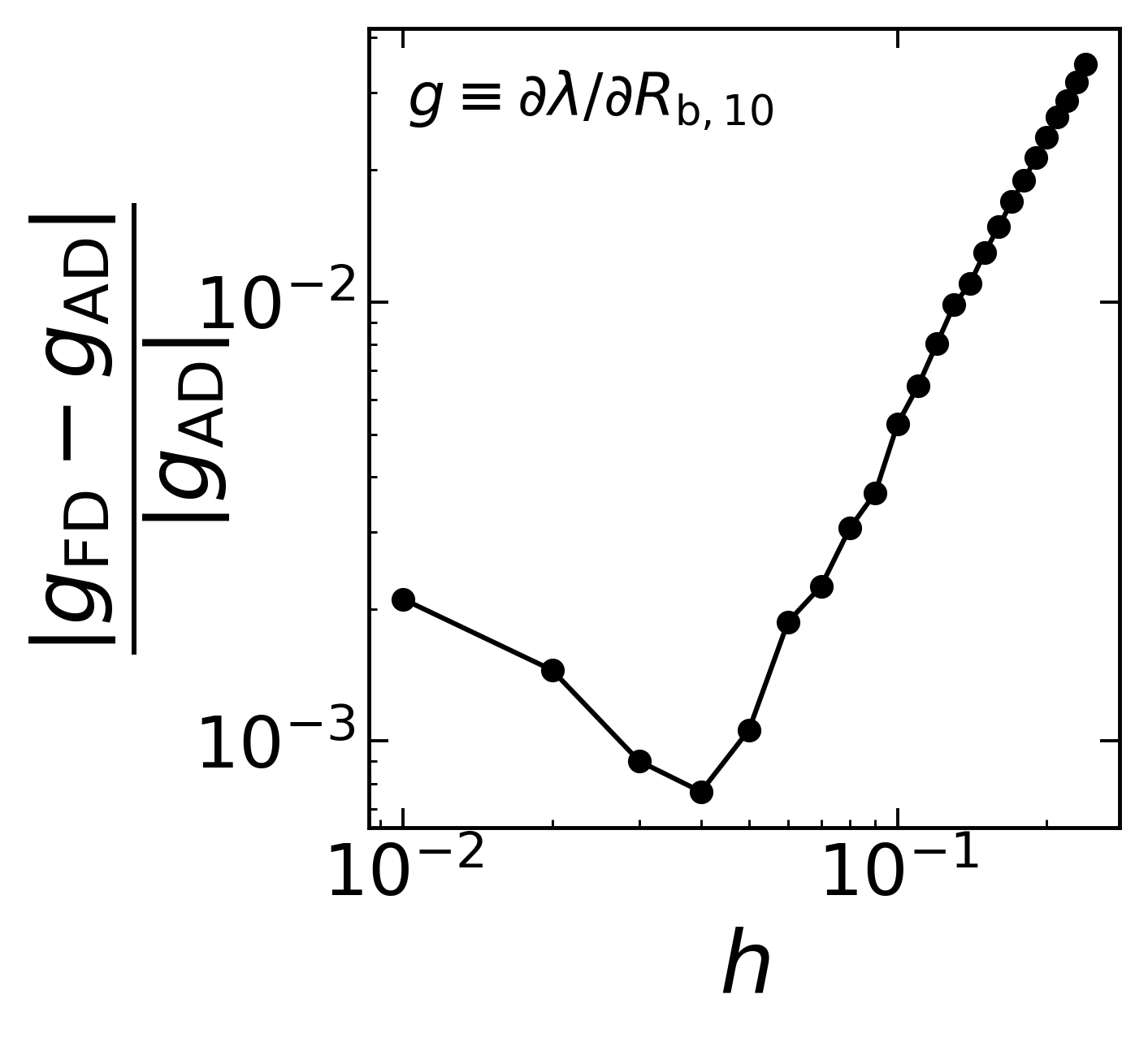}
        \caption{Rel. error with step size}
        \label{fig:FD-minus-AD}
    \end{subfigure}
\caption{We present the scan of the eigenvalue with respect to the $R_{\mathrm{b}, 10}$ boundary mode in figure~\ref{fig:landscape}, a comparison between the gradient $\partial \lambda/\partial R_{\mathrm{b}, 10}(R_{\mathrm{b}, 10} = 1.65)$ obtained from central FD with AD in figure~\ref{fig:FD-vs-AD}, and the scaling of the relative error between the two different gradient calculations as a function of the finite-difference step size $h = \Delta R_{\rm{b}, 10}$ in figure~\ref{fig:FD-minus-AD}.}
\label{fig:gradient-comparison}
\end{figure}

In this section, we demonstrate the ability to calculate the derivative of an unstable eigenvalue with respect to the equilibrium parameters using automatic differentiation. We use a reverse-mode gradient algorithm that can calculate the gradient of the eigenvalue with respect to boundary shape or equilibrium profile parameters without re-solving the equilibrium. Adjoint and reverse-mode gradient-based methods have previously been used in the context of coil design~\cite{paul_adjoint_2018}, quasisymmetry~\cite{Nies_Paul_Hudson_Bhattacharjee_2022}, neoclassical transport~\cite{unalmis2026spectrally}, infinite-$n$ ideal MHD stability optimization~\cite{Gaur2025OmnigenousStability}, etc. The reverse-mode gradient of an eigenvalue is defined by the following formula\footnote{Note that in practice it is much more memory efficient to calculate the gradient using the matrix--vector product $\partial(\hat{\mathbb{A}} \bm{v})/\partial\bm{x}_{\mathrm{b}}$ instead of materializing the gradient matrix separately and then multiplying with the eigenfunction.} 
\begin{equation}
    \frac{\partial \lambda}{\partial \bm{\mathrm{x}}_b} = \frac{\Big\langle \bm{v} \Big|\frac{\partial{\hat{\mathbb{A}}}}{{\partial \bm{\mathrm{x}}_b}}\Big| \bm{v}\Big\rangle}{\langle \bm{v} |\bm{v}\rangle}. 
    \label{eqn:HF}
\end{equation}
The derivation of~\eqref{eqn:HF} is provided in the appendix B of Gaur \textit{et al.}~\cite{Gaur2023AdjointBallooning}.
To test the reverse-mode automatic differentiation gradient and understand the optimization landscape, we create a family of LBD-QH-like equilibria by varying the boundary shape parameter $R_{\rm{b}, 10}$ while keeping the rest of the boundary and profile parameters fixed. This is equivalent to changing the minor radius of the equilibrium. For each equilibrium, we calculate the most unstable eigenvalue and the reverse-mode gradient of the eigenvalue $\partial \lambda/\partial R_{\mathrm{b}, 10}$ while enforcing the force-balance constraint and fixing the rest of the equilibrium parameters. The automatic differentiation (AD) gradient is then compared with a central finite-difference (FD) gradient for the family of equilibria. The eigenvalue landscape and gradient comparisons are presented in figure~\ref{fig:gradient-comparison}.

We find that the AD gradient matches well with the central FD gradient for a wide range of shapes.
The relative error between the two gradients is smallest at an intermediate step size, $h \approx 0.04$, where it is below $0.001$. For larger $h$, the truncation error of the central difference dominates. For smaller $h$, numerical noise in the individual eigenvalue evaluations dominates. We also notice a gradual reduction in the eigenvalue as the cross-section reduces. This happens because for a fixed enclosed toroidal flux $\psi_{\rm{b}}$, a smaller cross-section creates a higher magnetic field, reducing the normalized plasma pressure $\beta$ and hence the instability drive. 
 
Since the goal of this section was to plot the objective landscape and demonstrate AD calculation, the equilibria used for the landscape plots were a lower-resolution version of the LBD QH equilibrium used in the rest of the analyses in this paper. This causes the eigenvalues to differ from the eigenvalues in the rest of the sections. In this case, reducing the equilibrium resolution increased the eigenvalues.

All the calculations presented until this point were performed with $\Gamma = 5/3$ and included compressible modes. Most codes typically solve for incompressible modes so it would be advantageous for~\texttt{AGNI} to solve and optimize against incompressible modes. In the next section, we present two different methods to find the most unstable incompressible eigenmode in~\texttt{AGNI}.
\section{Incompressible modes}
\label{sec:incompressibility}
The most virulent modes to reactor operation are assumed to be incompressible which makes incompressibility an important physical limit. Moreover, almost all the previous solvers used the incompressibility condition to their advantage because it reduces the size of the potential energy matrix~\eqref{eqn:potential-energy}, improving memory and speed of the stability solvers. The typical process to enforce incompressibility ($\bm{\nabla} \cdot \bm{\xi} = 0$) is to write one of the components of $\bm{\xi}$ in terms of the other two components. For solvers working in spectral space, this is typically accomplished by first decomposing the perturbation into a field-aligned basis $\bm{\xi} = (\bm{\xi} \cdot \bm{b}_0/{B_0} )\bm{B}_0 + \bm{\xi}_{\perp}$ and then writing the incompressibility equation as
\begin{equation}
    \bm{B}_0\cdot \bm{\nabla} (\bm{\xi} \cdot \bm{b}_0/B_0) = -\bm{\nabla}_{\perp} \cdot \bm{\xi}_{\perp}
\end{equation}
In Fourier space, the operator $\sqrt{g} \bm{B}_0 \cdot \bm{\nabla}  \sum_{m, n} e^{i (m \theta + n \zeta)}=i \sum_{m, n} \psi^{'} (\iota m + n) e^{i (m \theta + n \zeta)}$, where $m$ and $n$ are poloidal and toroidal mode numbers, respectively. This allows one to algebraically replace the field-aligned component with the remaining components\footnote{For tokamaks, one can use the toroidal symmetry to simply replace the $\xi^{\zeta}$ component with $\xi^{\rho}, \upsilon$ by using the toroidal derivative $\partial/\partial \zeta = i n$ where $i$ is the imaginary unit and $n$ is the toroidal mode number.}. However, this is not currently straightforward to do in~\texttt{AGNI} as we perform all our calculations in real space.

Therefore, in this section, we will present two different numerical techniques to extract the most unstable incompressible eigenfunction. We will compare these techniques in terms of numerical complexity and accuracy, then demonstrate that the results from both the methods are consistent with each other, but the second method is less computationally expensive, making it more suitable for an optimization framework.
\subsection{Projection matrix method}
One solution is to use a projection matrix method to project out the compressible modes from the eigenvalue problem~\eqref{eqn:standard-EVP} by multiplying with a projection matrix
\begin{equation}
    \hat{\mathbb{P}} = \mathbb{I} - \hat{\mathbb{C}}^{T} (\mathbb{L}_G {\mathbb{L}_G}^{T})^{-1} \hat{\mathbb{C}},
\end{equation}
where $\mathbb{L}_G$ is the Cholesky decomposition of the Gram matrix $\mathbb{G} = \hat{\mathbb{C}} \hat{\mathbb{C}}^{T}$ and $\hat{\mathbb{C}}$ is the row-normalized compressibility operator such that $\hat{\mathbb{C}} \bm{v} = 0$, where the unnormalized operator $\mathbb{C} = \{C_{\rho}, C_{\theta}, C_{\zeta}\}$ comprises assembled compressibility matrices, given in~\eqref{eqn:compressibility-operators}. Note that $\hat{\mathbb{C}}$ has a shape $N \times 3N$,~\textit{i.e.}, it is not a square matrix and is applied to the displacement components ${\xi^{\rho_{\rm{s}}}, \xi^{\theta}, \xi^{\zeta}}$. The projector operator is multiplied by the reduced potential energy matrix $\hat{\mathbb{A}}$ to solve the modified eigenvalue problem
\begin{equation}
(\hat{\mathbb{P}}^{\dagger} \hat{\mathbb{A}} \hat{\mathbb{P}}) \bm{v} =   \lambda \mathbb{I}\bm{v}.
\end{equation}
The advantage of this method is that we can enforce incompressibility with a high accuracy $\bm{\nabla} \cdot \bm{\xi} \sim \mathcal{O}(10^{-8})$. The biggest disadvantage of this technique is that it increases the memory required ($\mathcal{O}(N^3)$) and takes longer due to the dense Cholesky transformation needed to obtain $\mathbb{L}_{G}$. Including the dense Cholesky transformation in the automatic differentiation loop makes optimization prohibitively expensive.

In the next subsection, we explain how we can satisfy incompressibility with a lower accuracy but in a way that is compatible with an optimization framework.
\subsection{Numerically increasing the speed of sound}
An efficient way to filter incompressible modes is to increase the penalty on the compressible modes by increasing the adiabatic constant $\Gamma$ in the potential energy~\eqref{eqn:potential-energy}. This will force the solver to find modes with reduced compressibility, i.e., $\bm{\nabla}\cdot \bm{\xi} \rightarrow 0$ as it finds the low energy state. An alternate way to think of this is as artificially increasing the speed of sound $c_{s} = \sqrt{\Gamma p/(M n_0)}$ which makes the fluid behave more like an incompressible fluid. This technique does not require a dense Cholesky calculation of a matrix like $\mathbb{L}_G$ and is compatible with an automatic differentiation framework.

\begin{figure}[h]
    \centering
    \begin{subfigure}[b]{0.35\textwidth}
    \centering
        \includegraphics[width=\textwidth, trim={2mm 2mm 1mm 1mm}, clip]{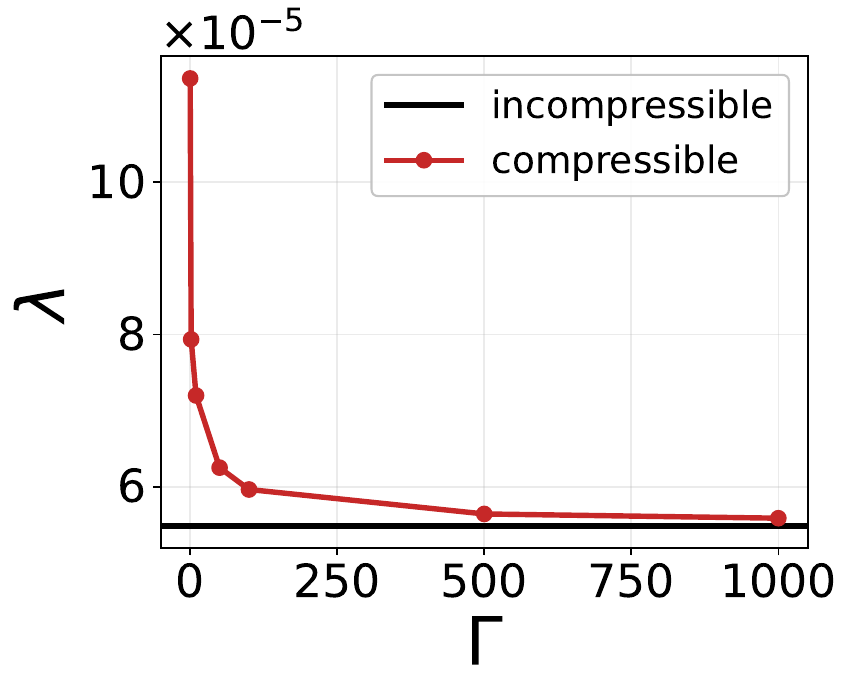}
        \vspace*{1mm}
        \caption{Eigenvalue comparison}
    \end{subfigure}
    \qquad
    \begin{subfigure}[b]{0.232\textwidth}
        \centering
        \includegraphics[width=\textwidth, trim={2mm 2mm 1mm 2mm}, clip]{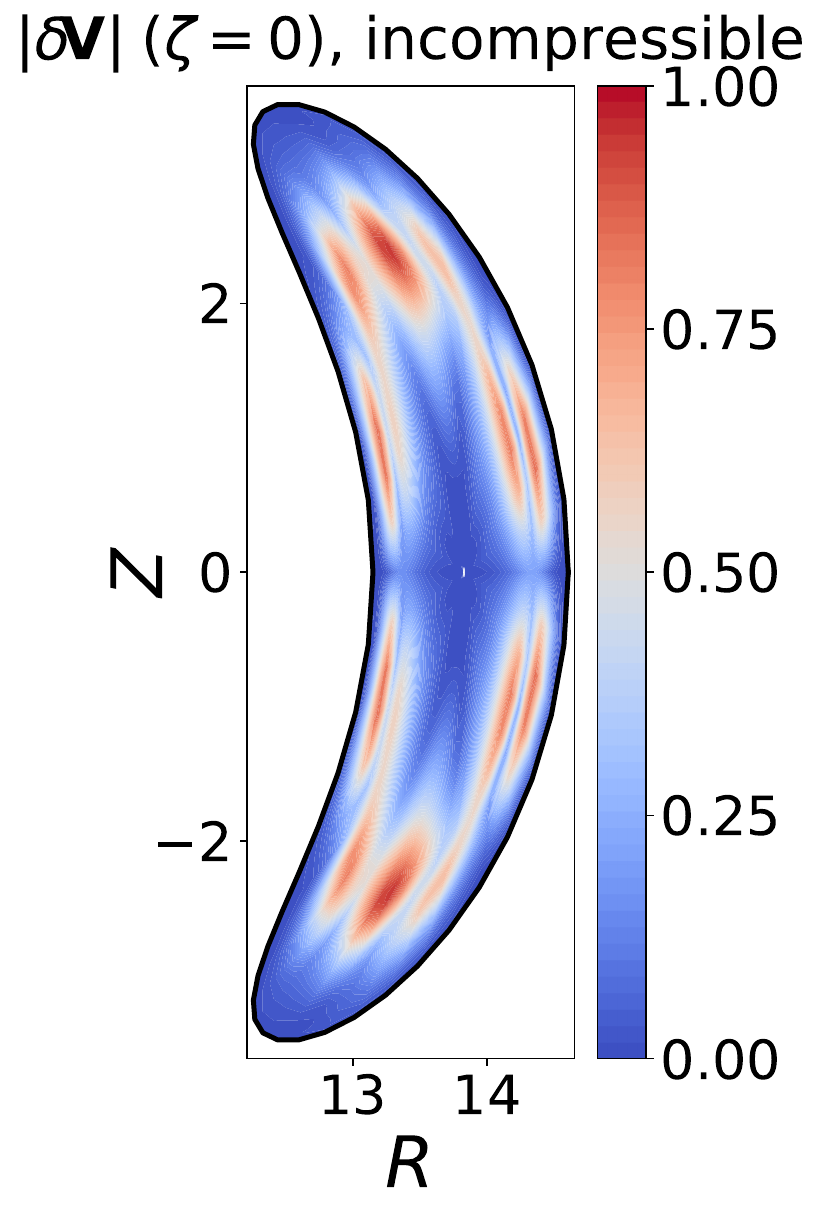}\\[0.8mm]
        \caption{Incompressible}
        \label{fig:Incompressible}
    \end{subfigure}
    \qquad
    \begin{subfigure}[b]{0.20\textwidth}
        \centering
        \includegraphics[width=\textwidth, trim={2mm 2mm 1mm 2mm}, clip]{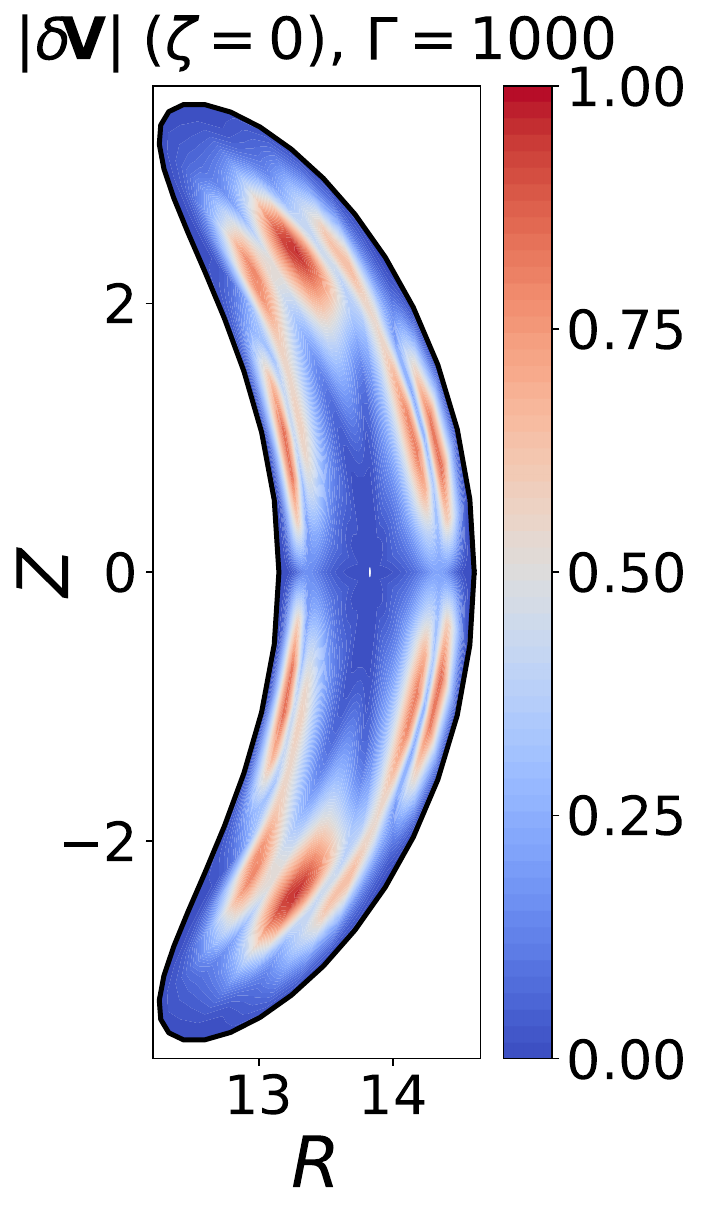}
        \caption{Compressible}
        \label{fig:Compressible}
    \end{subfigure}
\caption{Comparison of the incompressible and compressible eigenvalues and normalized eigenfunctions for the modified QH equilibrium. We observe the compressible eigenvalue approaches the incompressible limit as we increase the adiabatic constant $\Gamma$ and that the normalized incompressible and compressible ($\Gamma = 1000$) eigenfunctions agree closely.}
\label{fig:Incompressibility-test}
\end{figure}

To test the effectiveness of this method, we present a comparison of the two different ways of imposing incompressibility in figure~\ref{fig:Incompressibility-test}.

We observe reasonable agreement in the incompressible and compressible eigenvalues and eigenfunctions as the adiabatic constant increases. More importantly, we observe that compressible modes have a higher growth rate than incompressible modes and the most unstable compressible mode approaches the incompressible mode from the unstable side. This is numerically advantageous for the convergence of the solver because it provides a higher separation from the noise floor that we discussed in section~\ref{sec:numerical-methods}. Since both the methods agree well, we use the latter to impose incompressibility when performing optimization.

\section{Summary and conclusions}
\label{sec:conclusion}
In this manuscript, we presented~\texttt{AGNI} (Analysis of Global Normal modes in Ideal MHD), a differentiable ideal MHD stability solver and optimizer. In section~\ref{sec:MHD-equilibrium}, we described the ideal MHD model and the analytical form of the stability energy principle. In section~\ref{sec:discretization}, we described the discretization of the energy integral based on the differentiation matrix formalism in detail. In section~\ref{sec:numerical-methods}, we explained how the problem is simplified to a Hermitian eigenvalue problem, how we impose the boundary condition, and the numerical methods used to find the most unstable eigenmodes.
We also explored the practical limits on the eigenvalue resolution, and applied a custom differentiable eigendecomposition algorithm compatible with both CPUs and GPUs to find the most unstable mode. 
In section~\ref{sec:benchmarks}, we benchmarked~\texttt{AGNI} against~\texttt{NIMSTELL} for a strongly shaped quasi-helically symmetric stellarator, and a DSHAPE tokamak, and demonstrated reasonable agreement between both the codes. We ran the solver, performed a convergence study, and calculated evaluation times for the eigenvalues and its gradient with respect to the equilibrium parameters on both CPU and GPU. We demonstrated that for medium to high resolution problems, GPUs solve the eigenvalue problem up to an order of magnitude faster and perform gradient calculation around two orders of magnitude faster than CPUs, highlighting the advantage accelerated hardware provides us.

We also demonstrated the accuracy of the gradient calculated using automatic differentiation (AD) by comparing it with a gradient calculated with a finite-difference (FD) method and demonstrated good agreement. We observed that the AD gradient is robust even when the equilibrium is varied significantly. Finally, in section~\ref{sec:incompressibility}, we presented two ways of imposing incompressibility, demonstrated that they are consistent with each other, and showed that the method of numerically increasing the speed of sound by increasing $\Gamma$ gives us incompressible solutions, which are compatible with a gradient-based optimization framework.

There are three main directions in which this work will be extended. The first direction is to develop better preconditioners for the matrix-free version of~\texttt{AGNI}. This would allow us to reduce the optimization cost of high-resolution equilibria significantly. To fully ensure the stability of a plasma, the second goal is to include the effect of external (vacuum) modes which can destabilize the configuration even if the internal (plasma) modes are stable. This can be done using either the pseudo-vacuum approach~\cite{Gruber1985FiniteElementMHD, Anderson1990TERPSICHORE} or surface integral approach~\cite{merkel_free-boundary_1996, Avida2026OpenSourceEigenvalue}. The final feature would be to discretize the energy principle for ideal MHD equilibria with anisotropic pressure. This would allow us to study and optimize open configurations such as magnetic mirrors. 

\textbf{Acknowledgements:} R.G. would like to thank M. F. Martin, D. Dudt, A. H. Glasser, A. Bhattacharjee, C. Sun, F. Hindenlang, D. G. Panici, E. Kolemen, R. Conlin, Y. Zhou, E. Balkovic, B. Tripathi, X. Chu, M. Landreman, M. Lathrop, L. Lathrop, and C. Zhu for their feedback, and M. Avida for working on the vacuum stability calculation. 

\textbf{Funding:} This research used resources of the Perlmutter supercomputer located at the National Energy Research Scientific Computing Center (NERSC), a DOE Office of Science User Facility
supported by the Office of Science of the U.S. Department of Energy under contract No. DE-AC02-05CH11231 using NERSC award
FES-ERCAP0035536 and contract no. DE-AC02-09CH11466.
NIMSTELL calculations were supported by U.S. Department of Energy contract DE-FG02-99ER54546 and NERSC award FES-ERCAP0036386.

\textbf{Data availability:} The code is currently publicly available as a pull request  in the~\texttt{DESC} repository and also as a standalone package \href{https://github.com/rahulgaur104/AGNI}{here}.

\textbf{Declaration:} The authors used A.I. tools to refactor and test~\texttt{AGNI}. The refactored code was thoroughly tested against the original code to ensure equivalence.

\appendix
\section{Bases used for differentiation matrices and quadrature}
\label{app:Bases}
We have implemented several spectral and non-spectral bases as part of~\texttt{AGNI} in which the user can generate differentiation matrices and calculate the potential energy integral. For closed configurations like tokamaks and stellarators, the Fourier basis is the natural choice to calculate poloidal and toroidal derivatives. However, for the radial derivatives we typically use a non-periodic basis such as Legendre-Lobatto, Legendre-Radau (Gauss-Jacobi-Radau with $\alpha=0, \beta=0$), Chebyshev-Radau (Gauss-Jacobi-Radau with $\alpha = \beta = -0.5$), Zernike\footnote{Zernike is a non-separable basis that is useful for representing quantities on a two-dimensional disc while ensuring on-axis regularity. It uses Jacobi polynomials in the radial direction and Fourier in the poloidal direction.} (mixed radial-poloidal basis)~\cite{zernike_diffraction_1934}, finite-difference, or B-spline. The default basis is Gauss-Jacobi-Radau with $\alpha = -0.35, \beta = -0.75$.
\subsection{Numerical derivatives using different bases}
\label{subsec:diffmatrices}
\begin{figure}[h]
    \centering
    \begin{subfigure}[b]{0.31\textwidth}
    \centering
        \includegraphics[width=\textwidth, trim={1mm 1mm 1mm 1mm}, clip]{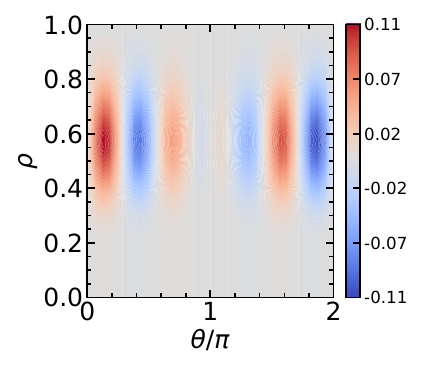}
        \caption{$f(\rho, \theta, \zeta = 0)$}
    \end{subfigure}
    \,
    \begin{subfigure}[b]{0.318\textwidth}
        \centering
        \includegraphics[width=\textwidth, trim={1mm 1mm 1mm 1mm}, clip]{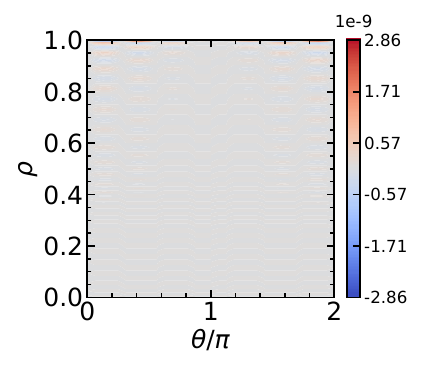}\\[0.2mm]
        \caption{GJR $(D_{\rho,0}f-\partial_{\rho}f)\lvert_{\zeta = 0}$}
        \label{subfig:derivative-error}
    \end{subfigure}
    \quad
    \begin{subfigure}[b]{0.32\textwidth}
        \centering
        \includegraphics[width=\textwidth, trim={1mm 1mm 1mm 1mm}, clip]{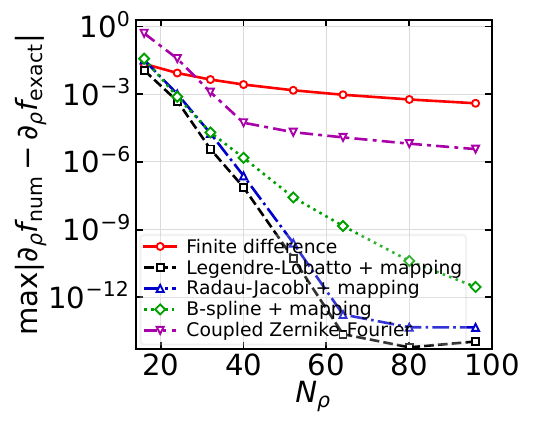}\\[1.5mm]
        \caption{Radial convergence scan}
    \end{subfigure}
\caption{Calculation of derivatives of the analytical function in~\eqref{eqn:analytical-f} and convergence with radial resolution for different bases. The grid points in the $\theta$ and $\zeta$ directions are held constant at $\nt = \nz = 32$ and the error in~\ref{subfig:derivative-error} is calculated at $\nr = 48$. The radial mapping function, defined in~\eqref{eqn:staircase-map} uses $m_1= 2, m_2 = 3, x_0 = 0.4$.}
\label{fig:numerical-derivatives}
\end{figure}

In this section, we will compare the derivatives calculated using differentiation matrices written in different bases. We use an analytical 3D function
\begin{equation}
    f(\rho, \theta, \zeta) = \rho^4 \exp(-20 (\rho-0.4)^2)(\sin(3 \theta) + \sin(4 \theta)) \cos(5 \zeta),
    \label{eqn:analytical-f}
\end{equation}
and present plots of the function and the convergence of derivatives comparing different bases in figure~\ref{fig:numerical-derivatives}.

We observe that the Legendre basis with the radial mapping function performs best out of all the bases but Gauss-Jacobi-Radau is more modular and performs almost as well as Legendre so it becomes the default radial basis in~\texttt{AGNI}. The radial mapping function allows the collocation nodes to move towards a specific $\rho$ from the ends, improving the efficiency of the derivative. The radial mapping function is defined in detail in the following section.

\subsection{Radial mapping function}
\label{subsec:radial-mapping-function}
To improve the effectiveness of the grid, we use a mapping function to distribute the collocation points so that they cluster where the most unstable eigenfunction peaks. For a mapping function $\rho_{\rm{s}} = f(\rho)$, we can write
\begin{equation}
\begin{gathered}
    D_{\rho_{\rm{s}}} = W_{\rm{s}}^{-1} D_{\rho}, \quad W_{\rm{s}} = \operatorname{diag}(f'(\rho)) \otimes \mathbb{I}_{\theta, 0} \otimes \mathbb{I}_{\zeta, 0} \\
\end{gathered}
\end{equation}
where the derivatives of $f$ are calculated using automatic differentiation in~\texttt{jax}. In practice, the most unstable eigenfunction peaks at a single $\rho$. Therefore, we choose 
\begin{equation}
\begin{split}
    f(\rho) = \epsilon + (1-\epsilon)&\bigg[ x_0\left(1 - e^{-m_1(\rho+1)} + 0.5 (\rho+1)e^{-2m_1}\right) \\
    &+ (1-x_0)\left(e^{m_2(\rho-1)} + 0.5 (\rho-1)e^{-2m_2}\right) \bigg]
\end{split}
\label{eqn:staircase-map}
\end{equation}
The coefficients $m_1$ and $m_2$ control the concentration of points from each end ($\epsilon$ and $1$) towards the point $\epsilon  \leq x_{0} \leq 1$. A higher value of $m_1$ will shift quadrature points from $\rho_{\mathrm{s}} = \epsilon$ towards $\rho_{\mathrm{s}} = x_0$ whereas a higher value $m_2$ will shift the points from $\rho_{\mathrm{s}} = 1$ towards $\rho_{\mathrm{s}} = x_0$.
The transformation is inspired by the Tal-Ezer-Kosloff~\cite{tal-ezer_accurate_1984} mapping but avoids the vanishing derivative $\lim_{\rho \rightarrow 1} f'(\rho) = 0$ near the ends. The exponential functions provide flexibility while avoiding a singularity.

\subsection{Sparsity structure for different bases}
\begin{figure}[h]
    \centering
    \begin{subfigure}[b]{0.2\textwidth}
    \centering
        \includegraphics[width=\textwidth, trim={2mm 2mm 1mm 2mm}, clip]{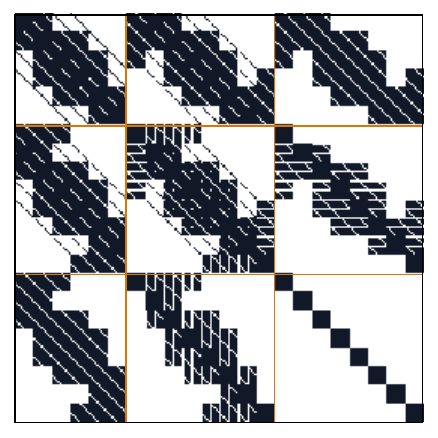}
        \caption{Finite-difference}
        \label{fig:Finite-diff}
    \end{subfigure}
    \qquad 
    \begin{subfigure}[b]{0.2\textwidth}
        \centering
        \includegraphics[width=\textwidth, trim={2mm 2mm 2mm 2mm}, clip]{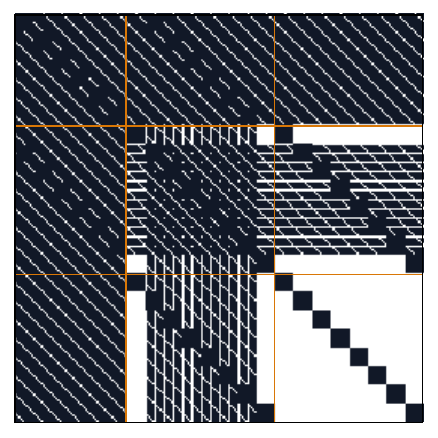}
        \caption{Gauss-Jacobi}
        \label{fig:Legendre}
    \end{subfigure}
    \qquad 
    \begin{subfigure}[b]{0.2\textwidth}
        \centering
        \includegraphics[width=\textwidth, trim={2mm 2mm 2mm 2mm}, clip]{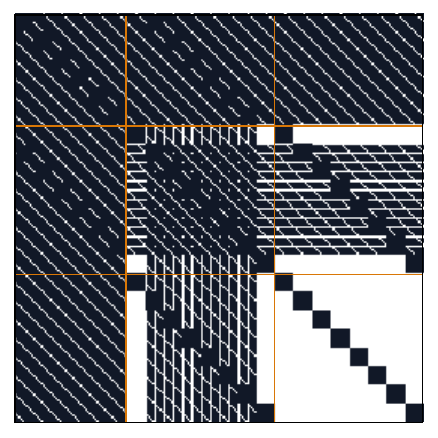}
        \caption{B-spline}
        \label{fig:B-spline}
    \end{subfigure}
    \qquad 
    \begin{subfigure}[b]{0.2\textwidth}
        \centering
        \includegraphics[width=\textwidth, trim={2mm 2mm 2mm 2mm}, clip]{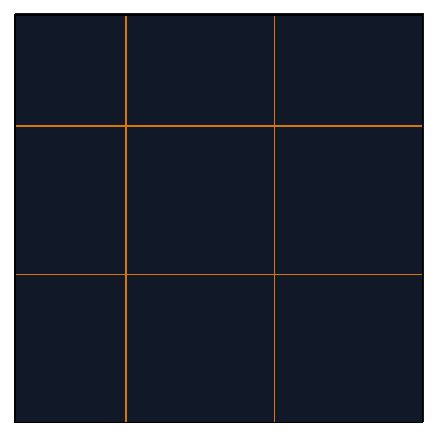}
        \caption{Zernike}
        \label{fig:Zernike}
    \end{subfigure}
\caption{Sparsity structure of the potential energy matrix for different choices of the radial basis. The toroidal and poloidal basis functions are Fourier. Every element in $\hat{\mathbb{A}}$ with an absolute value greater than $10^{-11}$ is a dark square. Note the absence of any sparsity in the potential energy matrix formed with the Zernike basis.}
\label{fig:Potential-energy-sparsity-structure}
\end{figure}

To understand the effect of different bases on the stability problem, we present sparsity plots of the final transformed matrix $\hat{\mathbb{A}}$ corresponding to different radial bases. We used a Fourier basis for the poloidal and toroidal directions but varied the radial basis. The sparsity structures are shown in figure~\ref{fig:Potential-energy-sparsity-structure}.
\label{subsec:sparsity-structure}

\section{Discretization of various terms in $\delta W$}
\label{subsec:discretization}
\label{app:Q2}
In this appendix, we will provide the discretized expressions of various terms in the energy integral in the differentiation matrix formulation.
\subsection{Fieldline bending term $\bm{Q}^2$}
Using the prescription provided in~\eqref{eqn:discretization}, we discretize the fieldline bending terms~\eqref{eqn:Q2_rho-rho}-\eqref{eqn:Q2_theta-zeta} that comprise $\bm{Q}^2$, as a part of the potential energy term
\begin{equation}
\begin{split}
    \int dV (\bm{Q}^2)_{\rho \rho} = {\xi^{\rho_{\rm{s}}}}^{\dagger}\Bigg[ &{D_{\theta}}^{\dagger} \frac{\psi^{'}}{\sqrt{g}} \iota^2 {\psi^{'}}^3 W  g_{\rho \rho}   D_{\theta} +  {D_{\zeta}}^{\dagger} \frac{\psi^{'}}{\sqrt{g}} W {\psi^{'}}^3 g_{\rho \rho} D_{\zeta} \\
    &+  \left({D_{\theta}}^{\dagger} \frac{\psi^{'}}{\sqrt{g}} \iota W {\psi^{'}}^3 g_{\rho \rho} D_{\zeta} + {D_{\zeta}}^{\dagger}  \frac{\psi^{'}}{\sqrt{g}} \iota W {\psi^{'}}^3 g_{\rho \rho} D_{\theta}\right) \Bigg] \xi^{\rho_{\rm{s}}},
\end{split}
\end{equation}
\begin{equation}
\begin{split}
    \int dV (\bm{Q}^2)_{\theta \theta} &= \upsilon^{\dagger}  \left( D_{\zeta}^{\dagger} \frac{\psi^{'}}{\sqrt{g}} W \psi^{'} g_{\theta \theta}  D_{\zeta} \right)\upsilon +  {\xi^{\rho_{\rm{s}}}}^{\dagger}  \Bigg[({D_{\rho}} \iota {\psi^{'}}^2 )^{\dagger} \frac{\psi^{'}}{\sqrt{g}} W  \frac{g_{\theta \theta}}{\psi^{'}} (D_{\rho} \iota {\psi^{'}}^2) \Bigg] {\xi^{\rho_{\rm{s}}}} \\
    &- {\xi^{\rho_{\rm{s}}}}^{\dagger} \left[ ({D_{\rho}}\iota {\psi^{'}}^2)^{\dagger} \frac{\psi^{'}}{\sqrt{g}} W g_{\theta \theta} {D_{\zeta}} \right] \upsilon + \rm{c.c.},\\
\end{split}
\end{equation}
\begin{equation}
\begin{split}
    \int dV (\bm{Q}^2)_{\zeta \zeta} &= \upsilon^{\dagger}  \left({D_{\theta}}^{\dagger} \frac{\psi^{'}}{\sqrt{g}} W \psi^{'} g_{\zeta \zeta} D_{\theta}\right)  \upsilon +  {\xi^{\rho_{\rm{s}}}}^{\dagger}  \Bigg[({D_{\rho}} {\psi^{'}}^2 )^{\dagger} \frac{\psi^{'}}{\sqrt{g}} W  \frac{g_{\zeta \zeta}}{\psi^{'}} (D_{\rho} {\psi^{'}}^2) \Bigg] {\xi^{\rho_{\rm{s}}}} \\
    &+ {\xi^{\rho_{\rm{s}}}}^{\dagger} \left[ ({D_{\rho}} {\psi^{'}}^2)^{\dagger} \frac{\psi^{'}}{\sqrt{g}} W g_{\zeta \zeta} {D_{\theta}} \right] \upsilon + \rm{c.c.},\\
\end{split}
\end{equation}
where the operators of the form $(D_{\rho}{\psi^{'}}^2)$ imply column-scaled differentiation matrices, and $\mathrm{c.c.}$ is used to denote the complex conjugate component corresponding to all the off-diagonal blocks, i.e., terms with different components of $\xi$. Next, we discretize the mixed (${\bm{Q}}^2_{xx'}$ s.t. $x \neq x'$) terms in $\bm{Q}^2$, 
\begin{equation}
\begin{split}
    \int dV (\bm{Q}^2)_{\rho \theta} &= {\xi^{\rho_{\rm{s}}}}^{\dagger} \left( {D_{\theta}}^{\dagger} \frac{{\psi^{'}}^3}{\sqrt{g}} \iota W g_{\rho \theta} D_{\zeta} +  {D_{\zeta}}^{\dagger} \frac{{\psi^{'}}^3}{\sqrt{g}}  W g_{\rho \theta} D_{\zeta} \right) \upsilon \\
    &- {\xi^{\rho_{\rm{s}}}}^{\dagger}\Bigg[ {D_{\theta}}^{\dagger}\frac{{\psi^{'}}^{2}}{\sqrt{g}} \iota W g_{\rho \theta} D_{\rho} \iota {\psi^{'}}^2  +  {D_{\zeta}}^{\dagger}\frac{{\psi^{'}}^2}{\sqrt{g}} W g_{\rho \theta} D_{\rho} \iota {\psi^{'}}^2 \Bigg]{\xi^{\rho_{\rm{s}}}},
\end{split}
\end{equation}
\begin{equation}
\begin{split}
    \int dV (\bm{Q}^2)_{\theta \zeta} &= -\upsilon^{\dagger} \left({D_{\zeta}}^{\dagger} \frac{\psi^{'}}{\sqrt{g}} W \psi^{'} g_{\theta \zeta}   D_{\theta}\right) \upsilon -{\upsilon}^{\dagger}\left( {D_{\zeta}}^{\dagger}\frac{\psi^{'}}{\sqrt{g}} W g_{\theta \zeta} {D_{\rho}}{\psi^{'}}^2  \right) \xi^{\rho_{\rm{s}}}\\
    &+ {\xi^{\rho_{\rm{s}}}}^{\dagger}\left[(D_{\rho} \iota {\psi^{'}}^2)^{\dagger} \frac{\psi^{'}}{\sqrt{g}} W  g_{\theta \zeta}  {D_{\theta}} \right] \upsilon  + {\xi^{\rho_{\rm{s}}}}^{\dagger}\Bigg[ (D_{\rho} \iota {\psi^{'}}^2)^{\dagger}  \frac{\psi^{'}}{\sqrt{g}} W \frac{g_{\theta \zeta}}{\psi^{'}} D_{\rho} {\psi^{'}}^2 \Bigg] \xi^{\rho_{\rm{s}}},
\end{split}
\end{equation}
\begin{equation}
\begin{split}
    \int dV (\bm{Q}^2)_{\rho \zeta} &= -{\xi^{\rho_{\rm{s}}}}^{\dagger} \left( {D_{\theta}}^{\dagger} \frac{{\psi^{'}}^3}{\sqrt{g}} \iota W g_{\rho \zeta} D_{\theta} +  {D_{\zeta}}^{\dagger} \frac{{\psi^{'}}^3}{\sqrt{g}}  W g_{\rho \zeta} D_{\theta} \right) \upsilon \\
    &- {\xi^{\rho_{\rm{s}}}}^{\dagger}\Bigg[{D_{\theta}}^{\dagger}\frac{{\psi^{'}}^{2}}{\sqrt{g}} \iota W g_{\rho \zeta} D_{\rho} {\psi^{'}}^2  +  {D_{\zeta}}^{\dagger}\frac{{\psi^{'}}^2}{\sqrt{g}} W g_{\rho \zeta} D_{\rho} {\psi^{'}}^2 \Bigg]{\xi^{\rho_{\rm{s}}}}.
\end{split}
\end{equation}

\subsection{The mixed term~\eqref{eqn:mixed}}
\begin{equation}
\begin{split}
  \int dV (\tilde{\bm{\xi}}\cdot \bm{\nabla}\rho)\left(\frac{\bm{j}_0 \times \bm{\nabla}\rho}{\lvert \bm{\nabla}\rho \rvert^2}\right)  \cdot \bm{Q} &\approx   {\xi^{\rho_{\rm{s}}\dagger}} \left[  {\psi^{'}}^{3} W \sqrt{g} \frac{(j^{\theta} g^{\zeta \rho} - j^{\zeta} g^{\theta \rho})}{g^{\rho\rho}} (\iota D_{\theta} + D_{\zeta}) \right] {\xi^{\rho_{\rm{s}}}} \\
  &+ {\xi^{\rho_{\rm{s}}\dagger}}  W {\psi^{'}}^2 \sqrt{g} \left(j^{\zeta} D_{\zeta} + j^{\theta} D_{\theta} \right)\upsilon\\
  &- {\xi^{\rho_{\rm{s}}\dagger}}  W {\psi^{'}} \sqrt{g} \left[ j^{\zeta} D_{\rho_{\rm{s}}}(\iota {\psi^{'}}^2 \xi^{\rho_{\rm{s}}}) - j^{\theta}D_{\rho_{\rm{s}}}({\psi^{'}}^2 \xi^{\rho_{\rm{s}}})\right].
  \label{eqn:mixed-discrete}
\end{split}
\end{equation}
Note that the mixed term will have a term which is the complex conjugate of the expression presented above. However, it is not shown here.
\subsection{Compressibility term}
Before discretizing the compressibility term, we define the compressibility derivative matrices
\begin{equation}
\begin{aligned}
    C_{\rho_{\rm{s}}} = D_{\rho_{\rm{s}}} + \frac{\partial \log(\sqrt{g})}{\partial \rho_{\rm{s}}} \mathbb{I},\\
    C_{\theta} = D_{\theta} + \frac{\partial \log(\sqrt{g})}{\partial \theta} \mathbb{I},\\
    C_{\zeta} = D_{\zeta} + \frac{\partial \log(\sqrt{g})}{\partial \zeta} \mathbb{I},
\end{aligned}
    \label{eqn:compressibility-operators}
\end{equation}
which allows us to write
\begin{equation}
\begin{split}
    \int dV \, \Gamma p_0 (\bm{\nabla} \cdot \tilde{\bm{\xi}})^2 &\approx {\xi^{\rho_s}}^{\dagger} ((C_{\rho_{s}}\psi^{'})^{\dagger}  \sqrt{g} W \Gamma p_0  (C_{\rho_{s}} \psi^{'}))\xi^{\rho_s} + (\upsilon + \xi^{\zeta})^{\dagger}(C_{\theta}^{\dagger} \sqrt{g}  W \Gamma p_0 C_{\theta}) (\upsilon + \xi^{\zeta})\\
    &+ {\xi^{\zeta}}^{\dagger} \left(C_{\zeta}^{\dagger} \frac{\sqrt{g} W \Gamma p_0}{\iota^2} C_{\zeta}\right) \xi^{\zeta} + {\xi^{\rho_s}}^{\dagger} ((C_{\rho_{s}} \psi^{'})^{\dagger} \sqrt{g} W \Gamma p_0 C_{\theta})(\upsilon + \xi^{\zeta}) \\
    &+ {\xi^{\rho_s}}^{\dagger} \left((C_{\rho_{s}} \psi^{'})^{\dagger} \frac{\sqrt{g} W \Gamma p_0}{\iota} C_{\zeta}\right)\xi^{\zeta}  + {(\upsilon + \xi^{\zeta})}^{\dagger} \left(C_{\theta}^{\dagger} \frac{\sqrt{g} W \Gamma p_0}{\iota} C_{\zeta}\right)\xi^{\zeta}+ \mathrm{c.c}.
\end{split}
\end{equation}

\pagebreak

\printbibliography

\end{document}